\documentclass[%
 aip,
 amsmath,amssymb,
reprint,
]{revtex4-2}

\usepackage{lineno}
\usepackage{mathtools}
\usepackage{graphicx}
\usepackage{booktabs,multirow}
\usepackage{braket}
\usepackage{textcomp}
\usepackage{yfonts}
\usepackage{amsfonts}
\usepackage[bbgreekl]{mathbbol}
\DeclareSymbolFontAlphabet{\mathbbm}{bbold}
\DeclareSymbolFontAlphabet{\mathbb}{AMSb}%
\usepackage{algorithm}
\usepackage{algpseudocode}
\usepackage[%
  bookmarks=true,
  colorlinks,
  linkcolor=blue,
  urlcolor=blue,
  citecolor=blue,
  plainpages=false,
  pdfpagelabels,
  final,
  breaklinks=true
]{hyperref}

\usepackage[normalem]{ulem}
\usepackage{xcolor}
\newcommand{\red}[1]{{#1}}

\begin{document}

\newcommand{\pd}[2]{\frac{\partial #1}{\partial #2}} 
\newcommand{\td}[2]{\frac{d #1}{d #2}} 

\newcommand{\bs}{\boldsymbol}
\newcommand{\bt}{\textbf}
\newcommand{\sech}{\text{sech}}
\newcommand{\erfc}{\text{erfc}}
\newcommand{\bse}{\begin{subequations}}
\newcommand{\ese}{\end{subequations}}
\newcommand{\im}{\text{i}}
\newcommand{\ud}[0]{\mathrm{d}}
\newcommand{\norm}[1]{\left\lVert#1\right\rVert}
\newcommand{\op}{\widehat}

\graphicspath{{Figures/}} 
\allowdisplaybreaks


\preprint{AIP/123-QED}

\title[Vectorial phase retrieval]{Vectorial phase retrieval in super-resolution 
polarization microscopy}

\author{R. Guti\'errez-Cuevas}
\email{rodrigo.gutierrez-cuevas@espci.fr}
 \affiliation{Institut Langevin, ESPCI Paris, Université PSL, CNRS, 
75005 Paris, France}
\affiliation{ 
Aix Marseille Univ, CNRS, Centrale Med, 
Institut Fresnel, Marseille 13013, France\looseness=-1
}
\author{L.A. Alem\'an-Casta\~neda}%
\affiliation{ 
Aix Marseille Univ, CNRS, Centrale Med, 
Institut Fresnel, Marseille 13013, France\looseness=-1
}%
\author{I. Herrera}
\affiliation{ 
Aix Marseille Univ, CNRS, Centrale Med, 
Institut Fresnel, Marseille 13013, France\looseness=-1
}
\author{S. Brasselet}
\affiliation{ 
Aix Marseille Univ, CNRS, Centrale Med, 
Institut Fresnel, Marseille 13013, France\looseness=-1
}
\author{M.A. Alonso}
\email{miguel.alonso@fresnel.fr}
\affiliation{ 
Aix Marseille Univ, CNRS, Centrale Med, 
Institut Fresnel, Marseille 13013, France\looseness=-1
}
\affiliation{The Institute of Optics, University of Rochester, 
Rochester, NY 14627, USA}

\begin{abstract} 
In single molecule orientation localization microscopy, valuable information about the orientation and longitudinal position of each molecule is often encoded in the shape of the point spread function (PSF). 
This shape, though, can be affected significantly by aberrations and other imperfections in the imaging system, leading to erroneous estimation of the measured parameters. 
A basic solution is to model the aberrations as a scalar mask in the pupil plane that is characterized through phase retrieval algorithms. 
However, this approach is not suitable for cases involving polarization-dependent aberrations, introduced either through unintentional anisotropy in the elements or by using birefringent masks for PSF shaping.
Here, this problem is addressed by introducing a fully vectorial model in which 
the polarization aberrations are represented via a spatially-dependent Jones matrix, commonly used to describe polarization-dependent elements. 
It is then shown that these aberrations can be characterized from a set of PSF measurements at varying focal planes and for various polarization projections. 
This \emph{PZ-stack} of PSFs, which contains both phase and polarization projection diversity, is used in a phase retrieval algorithm based on nonlinear optimization to determine the aberrations. 
This methodology is demonstrated with numerical simulations and experimental measurements. 
The \textsc{pyPSFstack} software developed for the modeling and
characterization is made freely available.
\end{abstract}

\maketitle

\section{Introduction}
Fluorescence microscopy is a widely-used imaging modality in biological research
\cite{Shashkova:2017} given its strong signal, selective labeling within complex
systems \cite{Dean:2014} and compatibility with super-resolution methods.\cite{Huang:2009} Moreover, this technique also allows access to the sample’s
structural properties \cite{Brasselet:2011,Brasselet:23}, making it very useful for studying
biomechanics at the molecular level. For example, in single-molecule orientation
localization microscopy (SMOLM), 3D spatial localization can reach a
precision of a few nanometers, while allowing  simultaneously the
characterization of the molecule's 3D orientational behavior (e.g.
mean orientation and degree of wobble).\cite{Brasselet:23} Common SMOLM  techniques include
polarization channel splitting \cite{Fourkas:2001,ValadesCruz:2016,
Caio:2022,Sison:2023} and point spread function (PSF) engineering,\cite{Backlund:2012,backer2014extending,Zhang:2022,curcio2020birefringent, Hullman:2021} which can
be used together or separately. The shape of the PSF can change considerably
with the emitter's orientation and longitudinal position, so it is crucial to
take this into account to enable a full estimation of the parameters
\cite{curcio2020birefringent,Ding:2021,Wu:2022} and to avoid localization biases.\cite{Enderlein:2006} In that respect, PSF engineering techniques have the aim of
enhancing these shape changes. Nevertheless, any optical aberration, polarization
distortion or misalignment in the imaging system can affect the final shape of
the PSFs and thus lead to inaccurate estimation of the parameters.
In particular,
polarization aberrations are delicate to correct for, as they require additional
adaptive strategies that account for the vectorial nature of light
propagation.\cite{Xia:2021,He:2022}

A common solution to this problem is to perform a set of calibration
measurements,\cite{hanser2003phase,hanser2004phase} and use them in a phase
retrieval optimization algorithm to determine the aberrations present in the
system. For this approach to work, it is important to have an accurate model for
light's propagation from a known source to the camera, and to incorporate phase
diversity, e.g. through measurements at varying focal planes, both to avoid falling
into local minima and to accelerate convergence.\cite{Gonsalves:2018} For
single-molecule localization microscopes, the standard approach is to measure
the PSFs generated by fluorescent nanobeads for varying focal planes and use
this Z-stack of images in a phase retrieval algorithm.\cite{Petrov:2017}
Initial approaches relied on scalar models assuming a point source emitting a
spherical wavefront along with a scalar pupil representing the aberrations. In
this simplified case, Gerchberg-Saxton iterative algorithms
\cite{gerchberg1972practical,fienup1982phase} can be used, thus reducing the
complexity of the implementation.  
However, these algorithms are less flexible since 
\red{the pupil model cannot be chosen freely, and they are not directly applicable to full vectorial models that consider the dipolar emission pattern and polarization distortions.\cite{hanser2003phase,hanser2004phase,clark2012microscope}}
A more flexible
approach is offered by casting the phase retrieval problem as a nonlinear
optimization routine where most parameters in the model can be found during the optimization process, although
this requires providing analytical formulas for the gradients in these
parameters, hence complicating implementation.\cite{fienup1982phase}
More accurate models that take into account the vector nature of the
emitted light have also been proposed.\cite{thao2019phase,ferdman2020vipr} They incorporate the effect of the interface between the medium embedding the
fluorescent particles and the coverslip, which causes extra aberrations,
polarization-dependent transmission and supercritical angle fluorescence (SAF)
radiation.\cite{novotny2006principles,axelrod2012fluorescence,ferdman2020vipr} However, these approaches 
assume a scalar mask to characterize all the remaining aberrations thus preventing them 
from correcting any residual polarization-dependent effects. Moreover, the 
polarized components of the emitted fields are eventually summed in
order to model an unpolarized measurement, which is not directly applicable to
polarized PSFs in SMOLM.

Recently, new SMOLM techniques have been proposed that use birefringent elements
either to encode efficiently the 3D orientation and 3D localization information
of the emitting dipole into the shape of the two polarization components of the
PSF,\cite{Backlund:2012,Zhang:2018,curcio2020birefringent,Zhang:2022} or to
understand the intensity and/or shape of their different polarization
projections.\cite{Caio:2022,Bruggeman:2023,Sison:2023} For such approaches, it
is essential to take into account the emission pattern of the dipole source, its
interaction with the interface between the embedded medium and coverslip, and
any polarization-dependent aberrations.\cite{hansen1988overcoming} To address these
issues, a model is used here where all vector aspects of the propagation of the
light emitted by the source to the back focal plane (BFP) are taken into
account, and where the aberrations and polarization distortions are represented by a birefringent
distribution at the pupil plane (BDPP) modeled with a spatially-varying Jones matrix.\cite{vella2018poincare} It is shown that in
order to properly characterize this BDPP, it is necessary to
introduce polarization projection diversity, obtained by projecting the PSFs onto various
polarization states, in addition to the phase diversity given by changing the
location of the focal plane. The PSF images generated with these two diversities
form a \emph{PZ-stack} that is fed into a nonlinear optimization algorithm that
allows retrieving the unknown BDPP. This approach allows including
many parameters that are necessary for a proper characterization, such as
photo-bleaching amplitudes, background illumination, or diversity-dependent
tilts.  
Accurate and computationally amenable models for the light produced by
fluorescent beads (commonly used for calibration measurements) are also
included. These models take into account the unpolarized nature of the emitted
light and the blurring due to bead size.\cite{alemancastaneda2022using} The
software package \textsc{pyPSFstack} used for the modeling of PZ-stacks and for
the phase retrieval process can be found in Ref.~\onlinecite{gutierrez2023repo}. The
retrieval algorithm was implemented with the neural network framework
\textsc{PyTorch},\cite{paszke2019pytorch} which greatly simplifies its
implementation and flexibility due to the automatic computation of all
gradients, and its integration with GPU if available.  
While the emphasis of this work is on the characterization of BDPPs, both the theory and code are equally applicable to scalar-only aberration
pupil distributions as shown in the  supplementary material (SM).

\section{Point-spread function of a dipolar source}

\subsection{Field at the back-focal plane}

\begin{figure*}
  \centering
  \includegraphics[width=.95\linewidth]{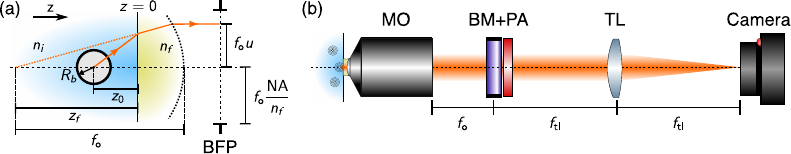}
  \caption{\label{fig:setup} Schematic of the experimental setup for the 
  collection and shaping of the emission by a source. 
  (a) Position of the fluorescent nanobead of radius $R_b$, 
  embedded in a medium of 
  refractive index $n_i$, with respect to the interface created by the 
  coverslip and the immersion liquid of the microscope objective, with
  refractive index $n_f$, and the focal plane located at $z_f$, 
  $z_f<0$ ($z_f>0$) if the focal plane is in the medium with refractive index 
  $n_i$ ($n_f$).
  (b) 
  Schematic of the collection arm composed of a microscope objective (MO), 
  followed by a birefringent mask (BM) and a polarization analyzer (PA) 
  at the back focal plane (BFP). 
  The light at the BFP is then focused onto the camera by the tube lens (TL)
  of focal length $f_\text{tl}$.}
\end{figure*}

In order to properly characterize a given system, one must first
derive an accurate model. Here, the situation depicted in Fig.~\ref{fig:setup}
is considered: the incoherent light emitted by a fluorescent bead is collected
by an immersion microscope objective with a high numerical aperture (NA). The
bead is assumed to be placed at an axial position $z_0$ from the interface between its
embedding medium with refractive index $n_i$ and the coverslip assumed to
have the same refractive index $n_f$ as that of the immersion liquid. (Because the bead is behind this interface, $z_0<0$.) The
index mismatch between these media introduces extra aberrations (see  SM
Secs. SI and SII for more details) and polarization-dependent transmission following the Fresnel
coefficients. It also allows the coupling of evanescent components emitted by
the bead when $n_f > n_i$ leading to SAF radiation, which can make up a
significant portion of the light detected by the camera.\cite{hellen1987fluorescence, axelrod2001total,axelrod2012fluorescence,axelrod2013evanescent}
All of these effects can be encapsulated into the Green tensor for a dipolar
source at the BFP of the microscope objective, which can be written as
\cite{Richards:1959,Wolf:1959,lieb2004single,novotny2006principles} 
\red{
\begin{align} \label{eq:Galphadelta}
  \mathbb G (\bt u) = & D^{(\Delta )}(\bt u) S (\bt u) \mathbbm g (\bt u) ,
\end{align}
where
\begin{align} 
  &S (\bt u) =  \exp \Bigg\{ 
  \im k n_f |z_0| \Bigg[ \frac{n_i}{n_f} 
  \sqrt{1 - \left(\frac{n_f u}{n_i} \right)^2 }
- \alpha \sqrt{1 - u^2} \Bigg]\Bigg\}, \nonumber \\
&D^{(\Delta )}(\bt u) =  \exp \left(
  \im k n_f \Delta  \sqrt{1 - u^2} \right)  , \label{eq:def}
\end{align}
and} $\mathbbm g$ is a $2\times 3$ matrix (see the SM Sec. SI for the explicit form)
that includes the effect of the Fresnel coefficients for the interface and
depends only on the normalized pupil coordinates at the BFP, $\bt u = (u_x,
u_y)$ whose maximum value is limited by the NA through $\norm{\bt u}_\text{max}
= \text{NA}/n_f$, 
and
$k=2 \pi/\lambda$ with $\lambda$ being the wavelength of the emitted light.
(Note that the definition of $\bt u$ differs from that in
Ref.~\onlinecite{alemancastaneda2022using} by a factor of $n_f$.) 
For simplicity, it
was assumed that the source is centered at the optical axis. The position of the
focal plane $z_f$ shown in Fig.~\ref{fig:setup} requires specifying two
parameters: $z_f = \Delta -\alpha |z_0|$, where $\alpha$ is a dimensionless
parameter fixing the position chosen for the reference focal plane (RFP) and
$\Delta$ the position  of the focal plane with respect to the RFP. The parameter
$\alpha$ is generally taken to be the one producing the best focus when
$\Delta=0$, although its definition is not unique  (one option being the one that minimizes the wavefront error\cite{Stallinga:2005}). 
As detailed in the
SM, minimizing the root-mean-square (rms) spot size with fourth-order
corrections for the wavefront difference between the SAF and defocus produces
the simple expression
\begin{align}
  \alpha = \frac{n_r^3 \left(32 n_r^2+11\right)}
  {24 n_r^4+16 n_r^2+3},
\end{align}
with $n_r = n_f/n_i$, which gives satisfactory results for both water/oil
($n_r=1.1391$) and air/oil ($n_r=1.515$) interfaces. This value is taken as the
default in this work, but it can be changed to match other experimental
configurations. In the SM Sec. SII we include other possible criteria for defining the best focus plane in our scenario: index matching between the objective immersion liquid and the coverslip. For situations in which this is not the case so that the coverslip has to be considered as a parallel slab of a finite thickness, expressions for the best focal plane can be found in Ref.~\onlinecite{Stallinga:2005}.

\subsection{Birefringence distribution at the pupil plane}

As mentioned in the introduction, the goal of this work is to characterize any residual aberrations and polarization-dependent effects due to stress and/or interfaces, the use of masks
aimed at tailoring the PSF, or a combination thereof.
These residual effects produce a birefringent distribution at the BFP, shown as a mask in Fig.~\ref{fig:setup},  which can be represented 
by a $2\times 2$ space-dependent 
Jones matrix,\cite{vella2018poincare}
\begin{align} \label{eq:jqw}
\mathbb J_\text{M}(\bt u) = e^{ \im 2\pi W(\bt u)} \left(
\begin{array}{cc}
q_0(\bt u) + \im q_3(\bt u) &q_2(\bt u) + \im q_1(\bt u) \\
-q_2(\bt u) + \im q_1(\bt u) & q_0(\bt u) - \im q_3(\bt u)
\end{array}
\right),
\end{align}
where $W$ represents a scalar aberration function, and the scalar pupil
functions $q_j$ are real. This matrix can be made unitary by enforcing the
condition $\sum_jq_j^2=1$, and otherwise it can include apodization effects.
The birefringence distribution $\mathbb{J}_{\rm M}$ can be used to represent
both a mask introduced to shape the PSFs, such as a stress-engineered optic
(SEO) \cite{spilman2007stress,spilman2007stressa,curcio2020birefringent} or a
q-plate,\cite{marrucci2006optical,rubano2019q,Zhang:2022} and the scalar and
polarization aberrations of the system. 
(The simplest case of a scalar mask 
corresponds to $q_0=1,q_1=q_2=q_3=0$.)
This Jones matrix acts on the Green tensor of the dipolar source at the 
BFP and the result is then propagated to the image plane via
\begin{align}
\mathbb G_\text{IP} (\tilde{\bs \rho}) = \iint  \mathbb J_\text{M}(\bt u)  
\cdot \mathbb G  (\bt u) 
\exp \left(- \im k  \frac{n_f \tilde{\bs \rho}}{M}  \cdot \bt u \right) \ud^2 u ,
\end{align}
where $\tilde{\bs \rho}$ denotes the transverse position at the 
image plane, and $M$ is the total magnification of the system. 
Note that for setups using a relay 
system the coordinates of $\mathbb G$ should be flipped, 
$\bt u \rightarrow -\bt u$.

For a fully polarized dipole oriented along the unit vector 
$\widehat{\bs \mu} = (\mu_x, \mu_y, \mu_z)$, the electric field 
distribution at the image plane 
is given by 
\begin{align}
\bt E_\text{IP} (\bt u) =
\mathbb G_\text{IP} (\tilde{\bs \rho} ) \cdot \widehat{\bs \mu}.
\end{align}
Therefore, the three columns of the Green tensor represent the field 
distribution at the pupil plane produced by a dipole along each of the three coordinate axes. 
Since the information about the orientation of
the dipole is encoded into the components of the Green tensor, in order to 
retrieve the dipole's orientation from the shape of the PSF
it is necessary to spatially separate its projections into two appropriately
chosen orthogonal polarization states, such as horizontal and vertical linear
or left and right circular. These polarization projections can be represented by
two matrices $\mathbb P_1$ and $\mathbb P_2$, thus for a fully polarized dipole
the PSF pair is given by
\begin{align} \label{eq:intprojcoh}
  I_{\text{IP},j} (\tilde{\bs \rho}) =  
  \norm{\mathbb P_j \cdot \mathbb G_\text{IP} (\tilde{\bs \rho})\cdot \widehat{\bs \mu}}^2 ,
\end{align}
with $j=1,2$. For unpolarized emitters such as fluorescent beads used for
characterization, the PSFs are given by the incoherent sum of the components of
the final Green tensor, which amounts to the incoherent sum of the dipoles
oriented along the three coordinates axes. In this case the pair of PSFs is
given by 
\begin{align} \label{eq:intproj}
  I_{\text{IP},j} (\tilde{\bs \rho}) =  
  \norm{\mathbb P_j \cdot \mathbb G_\text{IP} (\tilde{\bs \rho})}^2 .
\end{align}
Note that if no polarization projection is used the PSF is given by the sum of the pair of PSFs $I_{\text{IP},1}$ and $I_{\text{IP},2}$.

\section{Forward model for characterizing a birefringent distribution at the pupil plane}

\subsection{Polarization aberrations}

The goal of this work is to provide a method for determining from a set of calibration PSFs
the system's BDPP, 
represented by a Jones matrix of the form in Eq.~(\ref{eq:jqw}). The various
functions in this expression must first be expanded in terms of a basis whose
expansion coefficients are then determined through nonlinear optimization.
Common basis choices for optimization are the Zernike polynomials
\cite{yamamoto2007polarization} and a direct pixel representation,\cite{ferdman2020vipr} both of which result in comparable speeds
since they employ the same number of discrete Fourier transforms, the most costly operation even when implemented through the fast Fourier transform (FFT) algorithm. 

In what follows, a decomposition in the Zernike polynomial basis is used given
that its elements are easy to interpret 
and allow an accurate description with fewer parameters (although
examples using the pixel-based approach are shown in SM Sec. SVI). The
components of the Jones matrix are then decomposed as
\bse
\begin{align} \label{eq:wqs}
  W (\bt u) = &\sum'_l c_{l}^{(W)} Z_l(\bt u/u_\text{max}) ,
  \\
  q_{j} (\bt u) = &\sum_l c_{l}^{(j)} Z_l(\bt u/u_\text{max}),
\end{align}
\ese
where $j=0,\ldots,3$, and a single index notation was used for the basis
elements $Z_l$ (e.g. the Fringe notation\cite{Goodwin_2006}). 
Note that $\sum'$ in the expression for $W$ indicates that the terms
corresponding to piston and defocus are excluded.
The piston term would simply add a global phase that is unimportant and cannot be
determined from intensity measurements, while the defocus term would be redundant with
the more accurate defocus parameter $\Delta$ in Eq.~(\ref{eq:Galphadelta}) which
is also included as an optimization parameter. Also, note that any
misalignment of the PSFs with respect to the optical axis is corrected by the
scalar tilts present in $W$. The number of Zernike polynomials included in the decomposition depends on the expected spatial dependence of the BDPP. In this work, we found that the first 15 polynomials are usually sufficient, even when retrieving discontinuous BDPPs such as a q-plate (see Fig.~\ref{fig:seo_ret}). 
It should also be mentioned that the smooth description provided by the Zernike model works despite fast variations due to SAF radiation since these are already included in the propagation model. 
This Zernike expansion is inspired by the
Nijboer-Zernike theory where a scalar mask would be separated into real and imaginary parts before decomposing them in
terms of Zernike polynomials.\cite{janssen2002extended,braat2003extended,braat2005extended} 

\subsection{Phase and polarization diversity}

It is common practice in phase retrieval algorithms for optical microscopes to
assume access to a stack of intensity images for varying focal distances
$\Delta_{\zeta}$ from the location of the best focus (a Z-stack). The phase
diversity provided by the varying focal distances is taken into account by the phase factor $D^{(\Delta_\zeta)} (\bt u)$ defined in Eq.~(\ref{eq:def})
within the Green tensor in Eq.~(\ref{eq:Galphadelta}).
This additional information, referred to as phase diversity, helps the algorithm
converge to an appropriate solution without falling into local minima, as well
as discriminate between vortices with opposite topological charge. 

While a Z-stack is sufficient to determine scalar masks and aberrations, it is
not so for BDPPs since it does not discriminate between the true
$\mathbb J$ and its unitary transformations $\mathbb
J_{\mathbb U} = \mathbb U \cdot \mathbb J$, where $\mathbb U$ is a constant
unitary matrix, as exemplified in the following section. Therefore, it is
necessary to include information about the  polarization  dependence of the PSFs
used for the retrieval. This additional information is obtained by introducing a
polarization analyzer after the birefringent mask (see Fig.~\ref{fig:setup})
composed of a combination of waveplates and polarizers, where at least one
element rotates to generate various polarization projections of the output. This
polarization diversity is modeled by a set of constant Jones matrices $\mathbb
P^{(p)}$ that are applied to the Green tensor at the BFP along with the defocus
terms for the phase diversity in order to generate a PZ-stack of Green tensors
\begin{align}
  \mathbb G^{(\zeta ,p)}_\text{BFP} (\bt u) =  D^{(\zeta )} (\bt u) \mathbb P^{(p)} 
  \cdot \mathbb J_\text{M}(\bt u) \cdot \mathbb G_0  (\bt u).
\end{align}
It is important to notice that the constant matrices $\mathbb P^{(p)}$ cannot be 
unitary since they would give a unitary transformation  and thus 
have no effect of the shape of the PSFs. The 
simplest nonunitary matrix to implement is a projection matrix obtained 
by placing a linear polarizer at the end of the waveplate sequence.

\begin{figure*}
  \centering
  \includegraphics{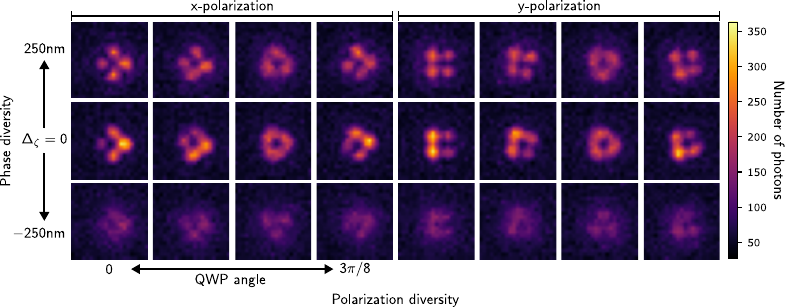}
  \caption{\label{fig:stack} PZ-stack for an unpolarized emitter 
  shaped with an SEO element and modeled with \textsc{pyPSFstack}. Only the PSFs
  at the initial, middle and final values of the defocus parameter
  $\Delta_\zeta$ used for the phase diversity in the retrieval of the BDPP
  shown in Fig.~\ref{fig:seo_ret} are shown. For each $\Delta_\zeta$ we show the
  PSFs for all polarization projections of the diversity. This polarization diversity
  is generated with a rotating quarter wave plate followed by a projection onto
  linear horizontal or vertical polarization states.
  }
\end{figure*}

This PZ-stack of Green tensors is then propagated to the image plane via
\begin{align}
  \mathbb G^{(\zeta ,p)}_\text{IP} (\bs \rho ) 
  = \iint \mathbb G^{(\zeta ,p)}_\text{BFP} (\bt u)  
 \exp \left( - \im k  \frac{n_f \bs \rho}{M}  \cdot \bt u \right) \ud^2 u ,
\end{align}
and its components are added incoherently (modeling an unpolarized source) to
obtain a PZ-stack of PSFs 
\begin{align}
  I^{(\zeta ,p)}_\text{IP} (\bs \rho ) = \norm{\mathbb G^{(\zeta ,p)}_\text{IP} }^2 
  = \sum_{i=x,y} \sum_{j=x,y,z} |G^{(\zeta ,p)}_{\text{IP},ij}( \bt u)|^2,
\end{align}
like the one shown in Fig.~\ref{fig:stack}. The polarization projections
$\mathbb P_1$ and $\mathbb P_2$ used to extract the orientation information can
also be used to define the polarization diversity as discussed in
Sec.~\ref{sec:num}. It is worth noting that while experimentally the
polarization diversity happens at the BFP, computationally it is better to
perform it at the image plane in order to avoid the computation of unnecessary FFTs. However, as discussed in Sec.~\ref{sec:exp},
there are some situations in which this is not possible. 

\subsection{Modeling the measured PSFs}
\label{Sec:ModelingPSF}
Before comparing the PZ-stack computed by the model presented thus far with one
measured experimentally, it is necessary to account for other effects. First,
depending on the size of the fluorescent bead it might be necessary to include a
blurring effect. As shown in Ref.~\onlinecite{alemancastaneda2022using}, the exact
three-dimensional blurring corresponds to a superposition of two-dimensional
convolutions between the PSFs generated by point sources located along the
longitudinal diameter of the bead with a kernel that weights more heavily the
contributions near the center of the bead and vanishes for those at the poles.
This exact blurring cannot be rewritten as a three-dimensional convolution due
to the lack of  translation invariance along the axis of propagation, but it
can be approximated through a semi-analytic method based on a Taylor expansion.
By keeping the first term of the expansion, we obtain a two-dimensional blurring
model based on a two-dimensional convolution of the PSFs. If instead we keep two
terms in the expansion we get a model for the blurring along both the transverse and longitudinal directions, which is
computed via two-dimensional convolutions of the PSFs and their
second-derivatives with respect to the bead's axial position thus giving a three-dimensional blurring model.  
Both of these approximate models are
computationally more efficient than the exact one and have been implemented in
\textsc{pyPSFstack}. Moreover, they can be used during the BDPP retrieval
process, as shown in Sec.~\ref{sec:blur}, albeit at a computational cost due to
the need of supplementary Fourier transforms. Nonetheless, they
allow for the analytic computation of the gradients, hence significantly limiting
the computational slowdown; using readily available functions to
blur images, on the other hand, such as the one used in Ref.~\onlinecite{ferdman2020vipr} with Gaussian kernel, would require the gradients to be computed via finite
differences.

The other two effects that must be considered are photobleaching of the
fluorescent beads and background illumination. Photobleaching causes the
number of photons emitted by the bead to diminish with time. Its effect can
be taken into account by implementing an overall amplitude factor
$a^{(p,\zeta)}$ that depends on both the phase and polarization diversities.
Background illumination is then added incoherently to the photobleached PSF
stack. The simplest model is to assume that the background illumination is
determined by a constant term $b^{(p,\zeta)}$ that depends on the phase and polarization diversities.
Extra terms for a spatially-dependent background can be added if needed.\cite{aristov2018zola} The modeled PZ-stack to be compared to the experimental
measurements is then given by 
\begin{align}
I^{(\zeta ,p)}_\text{tot} (\bs \rho )  
  =  a^{(p,\zeta)} \mathcal{B}\left[I^{(\zeta ,p)}_\text{IP} (\bs \rho ); R_b\right] 
  + b^{(p,\zeta)},
\end{align}
where $\mathcal{B}$ denotes the blurring operation that depends on the bead radius. In SM Sec. SIII we include a short derivation of the blurring operation when beads are modeled as solid spheres, following Ref.~\onlinecite{alemancastaneda2022using}.

\subsection{Assessing the accuracy with a cost function}

The last piece of the forward model to be considered 
is the choice of cost function 
used to tune the parameters so that the modeled PSFs, 
$I^{(\zeta ,p)}_\text{tot}$,
best fit the measured ones, $I^{(\zeta ,p)}_\text{exp}$.
In the absence of noise, any choice of cost function that has a minimum 
when the two quantities are the same should provide the same result. 
However,
noise is always present in experimental measurements and thus must be taken
account. In single molecule fluorescent microscopy one is normally limited by 
shot noise following a Poisson distribution, in which case the 
log-likelihood cost function,\cite{paxman1992joint}
\begin{align}
  \mathcal{C} = -\sum_{\zeta,p} \iint w(\bs \rho )\left\{ 
    I^{(\zeta ,p)}_\text{exp} (\bs \rho )
  \log \left[ I^{(\zeta ,p)}_\text{tot} (\bs \rho ) \right] 
  \right. \nonumber \\
  \left.
  - I^{(\zeta ,p)}_\text{tot} (\bs \rho )\right\} \ud^2 \rho,
\end{align}
should be used. Here, $w$ denotes a binary window function used to 
represent the region considered for the optimization due to a smaller size of the experimental data, and/or to exclude 
bad pixels of the camera. 
Another
common option for the cost function is the sum of differences squared,
which is appropriate when the noise follows a Gaussian distribution.
Both of these options are implemented in \textsc{pyPSFstack}.
Note that for the choice of cost function 
to be consistent, the values of $I^{(\zeta ,p)}_\text{exp}$ must 
actually follow the assumed distribution. 
This means that the images
should not be denoised and that the offset of the camera should be removed.

\section{Implementing the nonlinear optimization}
\label{sec:nlopt}

The goal of the nonlinear optimization routine is to find the set of
optimization parameters in the forward model that minimize the cost function
assessing the differences between the measured and modeled PZ-stacks. This is
achieved by supplying an optimization algorithm, such as Adam
\cite{kingma2017adam} or L-BFGS,\cite{liu1989limited} with a function that uses
all the current values of the parameters to compute the forward model all the
way to the value of the cost function. Additionally, one should supply the
optimization algorithm with another function that performs a backward
computation to obtain the gradients of the cost function with respect to all the
optimization parameters. These gradients are used to change the parameter values
until a minimum is reached. The gradient computation is straightforward but
tedious, and can be achieved by following the rules outlined in Ref.~\onlinecite{jurling2014applications}. However, an advantage of implementing the
nonlinear optimization with the neural network framework \textsc{PyTorch} is
that only the forward model must be implemented explicitly, since the backward model for computing the gradients of the various parameters is computed automatically. 
This framework also offers the most common optimization algorithms.

\begin{figure}
  \begin{minipage}{0.99\linewidth}
  \begin{algorithm}[H]
    \caption{Vectorial phase retrieval}\label{alg:vpr}
   \begin{flushleft}\textbf{Input} \\\end{flushleft}
  \hspace*{.2cm}
  \begin{tabular}{p{1cm}p{7cm}}
  $\bs \theta_\text{sys}$ & system parameters ($n_i$, $n_f$, $\lambda$, $z_0$, $\alpha$, $R_b$, $NA$, $M$, $p_\text{cam}$, $N$) \\
  $\bs \theta^{(0)}_\text{opt}$ & \emph{initial optimization parameters} ($\Delta$, $c_l^{(W)}$, $c_l^{(j)}$, $R_b$, $a^{(p,\zeta)}$, $b^{(p,\zeta)}$)\\
  $I^{(\zeta, p)}_\text{exp}$ & measured PSFs
  \\
  $\Delta_\zeta$ & $N_z$ defocuses for phase diversity\\
  $\mathbb P^{(p)}$& $N_p$ Jones matrices for polarization diversity\\
  $\mathcal B $ & blurring model\\
  $\eta$ & learning rate \\
  $N_\text{iter}$ & number of iterations\\
  \end{tabular}\\
  \begin{flushleft}\textbf{Output} \\\end{flushleft}
  \hspace*{.2cm}
  \begin{tabular}{p{1cm}p{7cm}}
  $\mathbb J_M$& Jones matrix at the pupil plane\\
  \end{tabular}\\
  \begin{flushleft}\textbf{Procedure} \end{flushleft}
  \hspace*{.2cm}
  \vspace*{-.5cm}
  \begin{algorithmic}[1]
    \State Initialize model with $\bs \theta_\text{sys}$, $\bs \theta^{(0)}_\text{opt}$, and $\mathcal B$
    \For{$n  \gets 1$ to $N_\text{iter}$}
    \State Compute forward model 
    \Comment{See Algorithm \ref{alg:forward}}
    \State Compute gradients
    \Comment{Uses automatic differentiation}  
    \State Update optimization parameters
    \Comment{Takes a step size of $\eta$ using the gradients}
    \EndFor
\end{algorithmic}
\end{algorithm}
\end{minipage}
\end{figure}

\begin{figure*}
\begin{minipage}{0.7\linewidth}
\begin{algorithm}[H]
  \caption{Implementation of the forward model}\label{alg:forward}
 \begin{flushleft}\textbf{Input} \\\end{flushleft}
\hspace*{.2cm}
\begin{tabular}{p{1cm}p{12cm}}
$\bs \theta_\text{sys}$ & system parameters ($n_i$, $n_f$, $\lambda$, $z_0$, $\alpha$, $NA$, $M$, $p_\text{cam}$, $N$) \\
$I^{(\zeta, p)}_\text{exp}$ & measured PSFs
\\
$\mathbbm g $ & Green tensor\\
$\Delta_\zeta$ & $N_z$ defocuses for phase diversity\\
$\mathbb P^{(p)}$& $N_p$ Jones matrices for polarization diversity\\
$\mathcal{B}$ & blurring model\\
$\bs \theta_\text{opt}$ & \emph{optimization parameters} ($\Delta$, $c_l^{(W)}$, $c_l^{(j)}$, $R_b$, $a^{(p,\zeta)}$, $b^{(p,\zeta)}$)
\end{tabular}\\
 \begin{flushleft}\textbf{Procedure} Forward model \end{flushleft}
\begin{algorithmic}[1]
\State $\mathbb G(\bs \ell) \gets  D^{(\Delta )}(\bs \ell) S (\bs \ell) \mathbbm g (\bs \ell)$
\Comment{Compute Green tensor} 
\State 
$\mathbb G_J(\bs \ell)  \gets  \mathbb J\left[\bs
\ell;c_l^{(W)},c_l^{(j)}\right] \cdot \mathbb G(\bs \ell) $ 
\Comment{Apply birefringent mask to Green tensor}
\State 
$\mathbb G^{(\zeta)}(\bs \ell)  \gets D^{(\Delta_\zeta)}(\bs \ell) \mathbb G_J(\bs \ell) $
\Comment{Apply phase diversity}
\State 
$\mathbb G^{(\zeta)}_\text{IP} (\bs \ell) \gets FFT \left\{ \mathbb G^{(\zeta)}(\bs \ell) \right\} $
\Comment{Propagate to image plane}
\State 
$\mathbb G^{(\zeta,p)}_\text{IP} (\bs \ell)  \gets \mathbb P^{(p)} \cdot \mathbb
G^{(\zeta)}_\text{IP} (\bs \ell)$ 
\Comment{Apply polarization diversity} 
\State
$I^{(\zeta,p)} (\bs \ell) \gets \sum_i \sum_j |G^{(\zeta ,p)}_{\text{IP},ij}(\bs \ell)|^2 $
\Comment{Compute intensity}
\State
$I^{(\zeta,p)}_\text{blur} (\bs \ell) \gets \mathcal{B}\left[ I^{(\zeta
,p)}_\text{IP}(\bs \ell), R_b \right] $ 
\Comment{Apply blurring} 
\State
$I^{(\zeta,p)}_\text{tot}(\bs \ell)  \gets a^{(p,\zeta)}
  I^{(\zeta,p)}_\text{blur} (\bs \ell) + b^{(p,\zeta)} $
\Comment{Photobleaching and background illumination}
\State 
$\mathcal{C}  \gets -\sum_{\bs \ell, \zeta, p} w (\bs \ell)\left\{ I^{(\zeta,
p)}_\text{exp}(\bs \ell) \ln \left[ I^{(\zeta, p)}_\text{tot}(\bs \ell)\right] -
I^{(\zeta, p)}_\text{tot} (\bs \ell)\right\} $ 
\Comment{Compute cost function}
\end{algorithmic}
\begin{flushleft}\textbf{Return} $\mathcal C$ \\\end{flushleft}  
\end{algorithm}
\end{minipage}
\end{figure*}

\red{
The implementation using \textsc{PyTorch} greatly simplifies the vectorial phase retrieval algorithm as shown in Algorithm~\ref{alg:vpr}.
First, the system parameters and the initial values for the optimization parameters are used to initialize the forward model used to compute the estimated PSFs and evaluate the current value of the cost function. 
For all the examples presented in this work, the BDPP was initialized as the identity matrix, i.e. $c_0^{(0)}=1$ and $c_l^{(W)}=c_l^{(j)}$ for all $j,l\ne 0$.  
However, this can be changed if there is \emph{a priori} knowledge about the BDPP. 
After initialization, the phase retrieval algorithm enters a For loop in which the forward model is used to compute the cost function using the current values of the parameters. 
Then, automatic differentiation is called to obtain the gradients. 
Finally, hyperparameters of the chosen optimization algorithm and the values of the gradients are used to update the values for all the optimization parameters. 
This process is iterated $N_\text{iter}$ times after which the process is completed.
Note that other criteria can be used to terminate the process: for example, the process could be terminated when the change of the cost function  falls below a set limit. 
All the results presented in this work were obtained using the Adam optimizer since it is more robust than simple gradient descent and faster and less memory intensive than L-BFGS. 
While this optimization algorithm has several parameters that can be tuned, in practice, it was found that choosing the appropriate learning rate was sufficient to obtain satisfactory solutions. 
The convergence was assessed by analyzing the evolution of the cost function.
A more detailed explanation can be found in the examples of the repository.\cite{gutierrez2023repo}
}

To implement the forward model, all calculations are performed  numerically
using FFTs, requiring all spatial quantities to be
sampled consistently with the camera's pixel pitch  $p_\text{cam}$, so that the
size at the BFP is fixed to $L_\text{pupil} = \lambda M/(p_\text{cam}n_f)$. The
total size at the BFP is divided into $N\times N$ points used for the computation, and
the resulting two-dimensional sampling is labeled with the lexicographic index
$\bs \ell$ which takes over all spatial dependence in terms of $\bt u$ and $\bs
\rho$, for instance $\mathbb G_0(\bt u) \rightarrow \mathbb G_0(\bs \ell)$. 
All the parameters and steps necessary to compute the forward model are summarized in Algorithm \ref{alg:forward}.

\begin{figure*}
  \centering
  \includegraphics[width=1\linewidth]{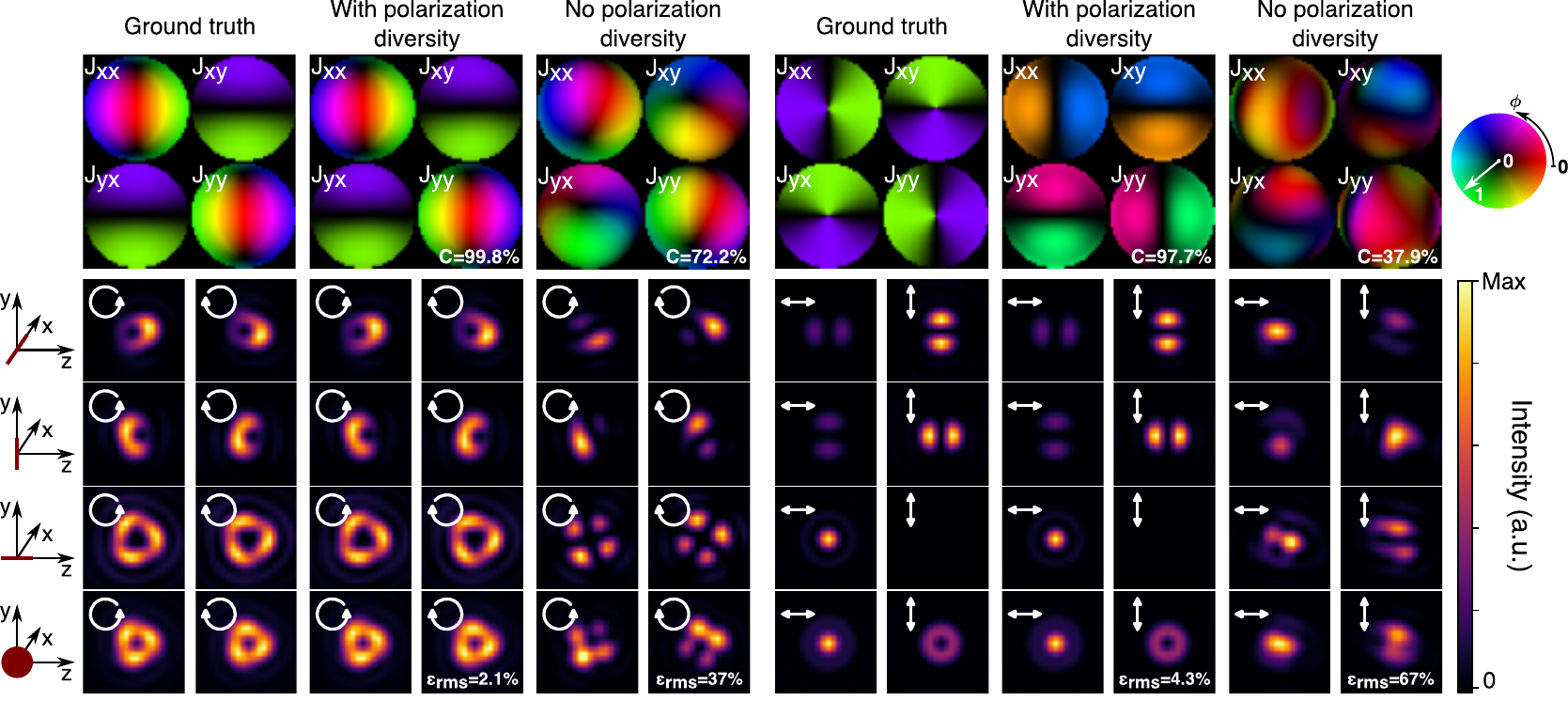}
  \caption{\label{fig:seo_ret} BDPP retrieval with and without 
  polarization diversity. (first row) Elements of the Jones matrix for the ground truth and the 
  retrieved BDPP with and without polarization diversity for (left) an SEO and (right) a q-plate. Also shown are the PSFs generated by point 
  dipoles (second to fourth row) oriented along each of the three Cartesian axes and 
  (last row) an unpolarized dipole, all for each of the standard projectors used 
  for each type of birenfringent window. For the SEO, the PSFs are projected onto left- 
  and right-circular polarizations, while for the q-plate they are projected onto linear horizontal and 
  vertical polarizations. Also given are the correlation between the retrieved and the true BDPP, as well as the RMS error  between the retrieved and true PSFs generated by point dipoles along each of the three Cartesian axes and an unpolarized dipole.}
\end{figure*}

\section{Numerical experiments} \label{sec:num}

\subsection{Polarization diversity VS more phase diversity}

To exemplify the implementation of the phase retrieval algorithm and, in
particular, the need for polarization diversity to properly characterize a
BDPP, we consider the retrieval of two masks used recently for
estimating the position and orientation of single emitters: an SEO\cite{spilman2007stress,spilman2007stressa}
\red{with
\begin{align}
  \mathbb J_\text{M} = 
  \cos \left(\frac{c u}{2}\right)
  \left( 
    \begin{array}{cc}
     1 & 0  \\
    0 & 1
    \end{array}
    \right)
    + \im \sin \left(\frac{c u}{2}\right) 
  \left( 
    \begin{array}{cc}
      \cos \phi & - \sin \phi  \\
    -  \sin \phi &  -  \cos \phi
    \end{array}
    \right),
\end{align}
and the parameter $c=1.25\pi$,
and a q-plate with unit topological charge,\cite{marrucci2006optical,rubano2019q,Zhang:2022} 
also known as a vortex waveplate, with
\begin{align}
  \mathbb J_\text{M} = \left( 
    \begin{array}{cc}
    \im \cos 2 \phi & \im \sin 2 \phi  \\
    \im \sin 2 \phi &  -\im \cos 2 \phi
    \end{array}
    \right),
\end{align}
both of which are shown 
}
in Fig.~\ref{fig:seo_ret}. Following the strategy outlined thus far,
\textsc{pyPSFstack} is used to model a PZ-stack for each BDPP such
as the one shown in Fig.~\ref{fig:stack} for the SEO. For the phase diversity,
images are taken from $-250$nm to $250$nm of the RFP with a step
size of 50nm. For the polarization diversity, a quarter wave plate (QWP) is
rotated from $0$ to $3\pi/8$ with a step of $\pi/8$ and is followed by a
Wollaston prism that projects the output onto horizontal and vertical linear
polarizations. This choice is inspired by the setup used in Ref.~\onlinecite{curcio2020birefringent} where the SEO is followed by a QWP and a Wollaston
prism to project the PSF into left and right circular polarizations. It is also
assumed that 10000 photons arrive on average to the camera to form the PSFs, to
which an additional 50 photons per pixel are added as background. Noise
following a Poisson distribution is incorporated. 
\red{Note that the choice on diversity follows the results presented in the SM Sec. SV, where the accuracy of the retrieval is studied as a function of measured photons for various combinations of phase and polarization diversity.}
For simplicity, we consider a small fluorescent bead with $R_b=10\text{nm}$ so
that spatial blurring is negligible. Nonetheless, the exact blurring model
introduced in Ref.~\onlinecite{alemancastaneda2022using} was used to compute the PZ-stacks
for testing the BDPP retrieval algorithm. Moreover, a random error
of the order of 20nm was introduced to the distance between the bead and the
coverslip, and one of the order of 25nm to the location of the RFP. 
\red{The algorithm was run on a desktop computer with an AMD Ryzen 9 3900X 12-Core CPU, and a NVIDIA GeForce RTX 2080 GPU.
Using a pupil sampling of $128 \times 128$ points the full retrieval routine consisting of 200 iterations took only a few seconds (see Table \ref{tab:run} for more details).
}

\setlength{\tabcolsep}{8pt}
\begin{table}
\caption{Runtime per iteration for a pupil sampling of $128 \times 128$ points, eleven phase diversities and eight polarization diversities for different blurring models and with and without the diversity-dependent tilts mentioned in Sec.\ref{sec:exp}.}
\label{tab:run}
\centering
\begin{tabular}{p{4.2cm} c  c } 
 \hline\hline
\multirow{2}{*}{Model} & \multicolumn{2}{c}{Time per iteration}\\
  & CPU &  GPU \\ [0.5ex] 
 \hline
 No blurring and no tilts &45 ms &9.6 ms \\ 
 2D blurring and no tilts& 110 ms & 13.7 ms   \\ 
 3D blurring and no tilts&  273 ms &21.4 ms   \\ 
 No blurring and with tilts  & 222 ms & 16.7 ms \\ 
 2D blurring and with tilts  & 287 ms & 20.6 ms \\ [1ex] 
 \hline\hline
\end{tabular}
\end{table}

\begin{figure*}
  \centering
  \includegraphics[width=.7\linewidth]{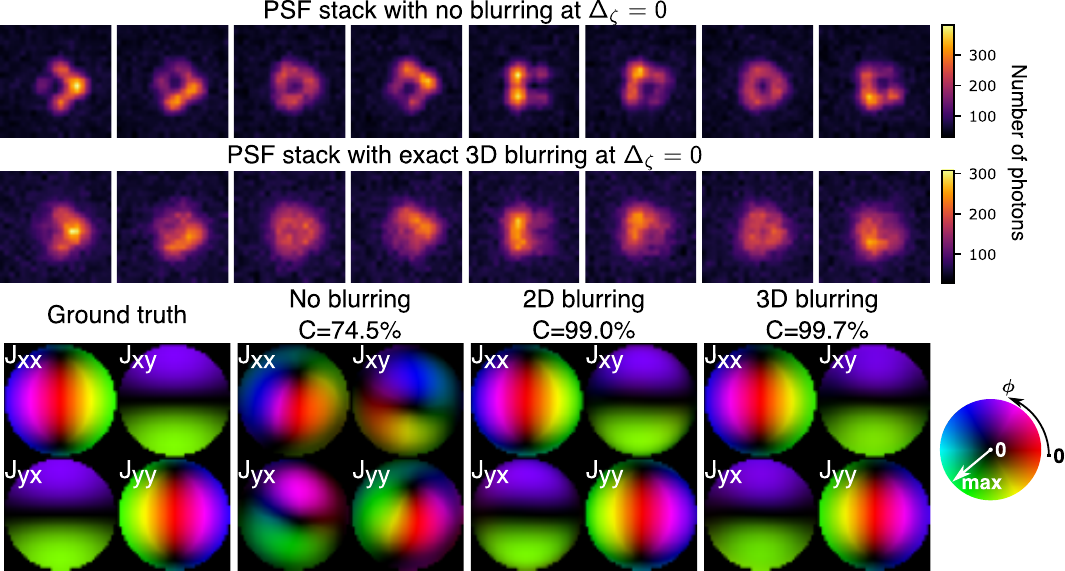}
  \caption{\label{fig:blur}BDPP retrieval from highly blurred PSFs. 
  Comparison of PSFs for an SEO at the reference focal plane ($\Delta_\zeta=0$) 
  for the same polarization diversities as those shown in Fig.~\ref{fig:stack} for 
  (first row) a point source and (second row) a fluorescent bead with radius 
  $R_b=150$ nm. (third row) Elements of the Jones matrix for the ground truth, and the retrieved 
  BDPPs without blurring, with a two-dimensional blurring model, 
  and a three-dimensional semi-analytic blurring model.
  }
\end{figure*}

The results of the procedure are shown in Fig.~\ref{fig:seo_ret}, where the
Jones matrix for the ground truth is compared to the ones retrieved from the PZ-stacks. 
This figure also shows the
PSFs formed by dipoles oriented along the three Cartesian axes and by an unpolarized dipole (constructed as an incoherent mixture of the three orthogonal dipoles). 
The PSFs shown are modeled using the standard projectors $\mathbb P_1$ and $\mathbb P_2$ (Eqs.~(\ref{eq:intprojcoh}) and (\ref{eq:intproj})) for each 
birefringent mask: for the SEO the output is projected onto 
left and right circular polarizations, while 
for the q-plate the output is projected onto the horizontal and vertical 
polarizations. 
\red{To assess the accuracy of the retrieval, both the correlation between the retrieved and true BDPPs as well as the root-mean square (RMS) error, $\epsilon_\text{RMS}$, between the retrieved and true PSFs are shown in Fig.~\ref{fig:seo_ret}.}
For the SEO, the retrieved BDPP and 
the corresponding PSFs are almost indistinguishable from those corresponding to the ground truth. 
However, for the 
q-plate there are appreciable differences between the original and retrieved BDPPs. 
The deviation at the center is due to the chosen model, since the Zernike
polynomials struggle to reproduce the singularity at the center of the q-plate. 
\red{
In general, this type of singularity does not arise as an aberration, but rather by design of the mask, so that it can be incorporated as prior information. For instance, on top of the Zernike decomposition, a q-plate could be added to the model with its center and orientation as optimization parameters.}
Another noticeable difference for the case of the q-plate is that the phase
between the two rows of the Jones matrix is not correct. 
This error happens only for birefringent masks that generate rotationally symmetric PSFs
for unpolarized emitters
for all polarization projections. For these masks, the 
algorithm cannot determine the global phase between the rows of the corresponding Jones 
matrix. 
This problem can be solved by placing the device that introduces the polarization diversity before the birefringent 
mask. 
Even when this solution is not used, the correlation (when neglecting the phase difference between the rows) is quite high and the reproduced PSFs for any dipole orientation are again indistinguishable 
from the true ones. 
Therefore, these differences are inconsequential for the purposes 
of single molecule fluorescence microscopy.
Additionally, the error introduced for $z_0$ and $\alpha$ has no impact
on the retrieval as long as the defocus $\Delta$ is included as an
optimization parameter. 
\red{Similar results using a pixel-based model are shown in the SM Sec. VI, as is a comparison of the convergence of the cost function between the two models.}

For comparison, the retrieval for the same birefringent mask is also performed without using 
polarization diversity. To make this comparison consistent and fair, the number of photons 
reaching the camera and those in the background illumination are doubled, since there is no 
polarization separation. 
Moreover, the phase diversity images are now 
taken from $-250$nm to $250$nm of the 
RFP with a reduced step size of 5nm so that the total number
of PSFs used is larger than in the previous case. The results are shown in 
Fig.~\ref{fig:seo_ret}, where the algorithm is seen to fail to retrieve the 
appropriate BDPP, and thus it cannot generate the correct PSFs when they are projected 
onto a given polarization state. These results show the need for polarization 
diversity when characterizing a BDPP. 
\red{A more detailed comparison between the scalar and vector models is provided in the SM Sec. IV. Note that runtime is much longer  due to the increase in the number of FFTs required. (This runtime should be comparable with that of a model that includes tilts but no blurring; see Table \ref{tab:run}).}

\subsection{Incorporating blurring due to size to the birefringent distribution retrieval}
\label{sec:blur}

As an additional test of the present implementation, the retrieval of a BDPP from a highly blurred PZ-stack like the one shown in Fig.~\ref{fig:blur} is 
considered. Using the SEO as an example, a PZ-stack is constructed as in the previous 
section with the significant difference of increasing the radius of the nanobead, which 
is chosen randomly from a uniform distribution between 150nm and 170nm. Here again the 
distance to the coverslip is taken to be equal to the radius of the bead. As shown in Fig.~\ref{fig:blur}, this bead size causes a significant blur in the resulting PSFs. The blurred PSFs were computed with the exact 
blurring presented in Ref.~\onlinecite{alemancastaneda2022using}, but this approach turns out to be 
computationally expensive 
and not necessary for the retrieval procedure as shown in what follows.

A first attempt to retrieve the BDPP is performed by completely neglecting the blurring 
effect. However, as shown in Fig.~\ref{fig:blur}, this approach fails to retrieve the 
appropriate BDPP, showing that bead size effects cannot be neglected in this case. Next, as discussed in the SM Sec. SIII, the two approximate models presented in Ref.~\onlinecite{alemancastaneda2022using} were used for the 
retrieval where the radius of the bead is used as an 
optimization parameter for the blurring with an initial value of 150nm. The first of these approaches is a
2D convolution with a kernel given by a spherical Bessel 
function. This approach produces a satisfactory result, whose retrieved BDPP 
has a correlation of 98.6\% with the true one. 
The second is the semi-analytic model described in Section~\ref{Sec:ModelingPSF} for
reproducing the effects of the three-dimensional blurring, based on a 
Taylor expansion around the center of the bead.
In this case, a BDPP 
indistinguishable from the true one is again retrieved with a slightly better 
correlation of 99.5\%.
\red{
As shown in Table~\ref{tab:run}, both blurring models significantly increase the time per iteration due to the required supplementary Fourier transforms. This increase is particularly important for the 3D blurring model, which requires the computation of the first and second derivatives of the Green tensor with respect to the bead's axial position, hence tripling the number of Fourier transforms, in addition to two convolutions at the image plane.
Therefore, the small gain provided by the 3D model might not justify the extra computational resources needed for it.
Nonetheless, both models are able to perform the full retrieval in well under a minute with the CPU and in a few seconds with the GPU.}

\section{Characterization from experimental data}
\label{sec:exp}

\begin{figure*}
  \centering
  \includegraphics[width=.9\linewidth]{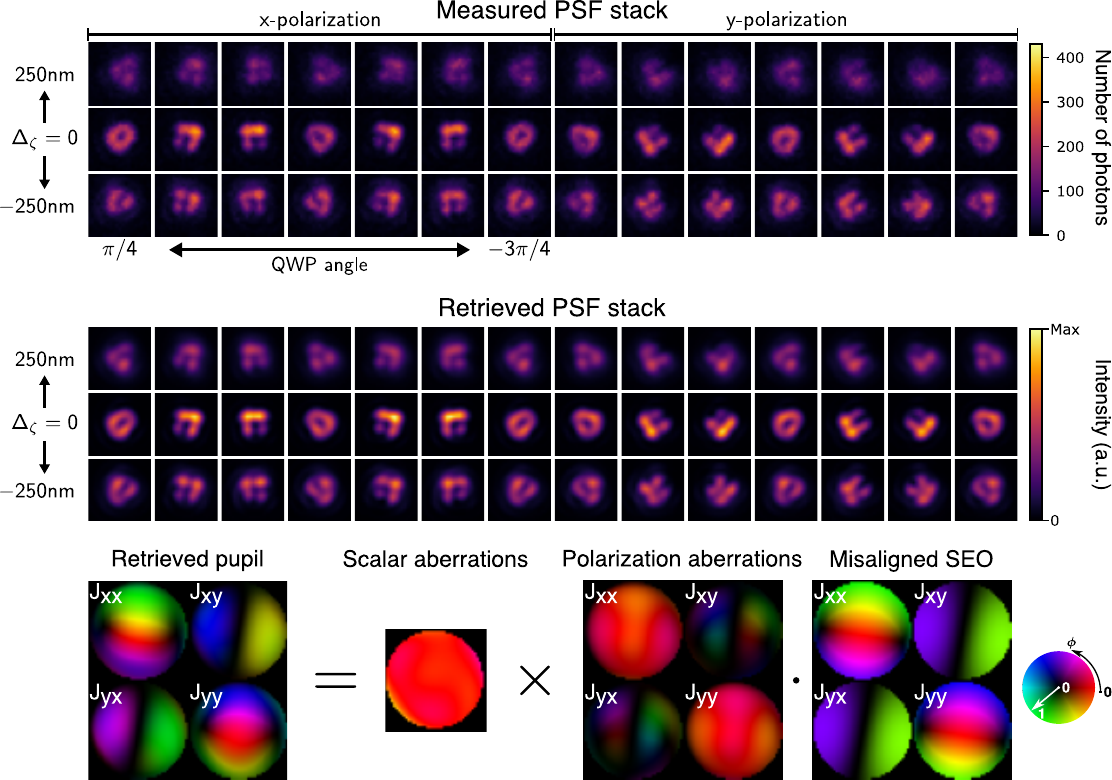}
  \caption{\label{fig:exp} BDPP retrieval from experimental data, taken from fluorescent nanobeads of diameter 20nm. (first row) Measured and (second row) retrieved PZ-stacks, where only the PSFs
at the initial, middle and final values of the defocus parameter $\Delta$ used for the phase
diversity are shown, for all polarization projections comprising the polarization diversity. (last row) 
The retrieved elements of the Jones matrix for the BDPP and its decomposition into a misaligned SEO element, 
and scalar and polarization aberrations. }
\end{figure*}

After validating the proposed retrieval procedure on simulated data, we apply it to retrieve a BDPP distribution from a PZ-stack measured experimentally. 
We use a sparse sample of fluorescent nanobeads with a diameter of 20 nm  
(orange carboxylate-modified FluoSpheres), immobilized on the surface of a 
poly-L-lysine-coated coverslip and embedded in water.  The sample is mounted 
on a XYZ piezo stage (Physik Instrumente), and is excited by a continuous wave 
laser emitting at 561 nm (Oxxius L4Cc) in a wide-field illumination 
configuration using an oil immersion objective lens 
(APO TIRF $\times$100, NA$=1.49$, Nikon).  The emitted fluorescence 
is collected by the same objective lens, and then passes through two multiband dichroics (Semrock, R405/488/561/635-t1-25x36) and a
fluorescence filter (Semrock, 605/40). A telescopic relay system (composed of two achromatic 
doublets with $f=250$ mm so that the magnification is unity) is used to access the BFP of the 
objective. The different polarization projections are taken by using a QWP (AQWP05M-600, Thorlabs) 
placed on a rotating mount (Newport, PR50CC), followed by a quartz Wollaston 
polarizing 2.2° beamsplitter (Edmunds, 68-820). The final images are 
measured using an ORCA Fusion-Digital CMOS C14440-20 UP (1024$\times$1024 pixels, 
6.5 $\times$ 6.5 $\mu$m pixel size, Hamamatsu). PZ-stacks were acquired with a 
step size of $50$ nm, and a rotation step for the QWP of 30$^\circ$. Figure \ref{fig:exp} 
shows part of the experimental PZ-stack.

Before launching the retrieval on the experimental PZ-stack, it should be 
noted that there are several factors that lead to the introduction of a 
diversity-dependent phase tilt at the BFP. First, the use of a Wollaston 
prism to separate spatially the two polarization components onto different 
sections of the camera might make it difficult to have the same center for the 
PSFs for each polarization component. Second, any slight wedge on the 
rotating QWP introduces a tilt that rotates with it. Finally, 
any slight misalignment between the stage moving the sample and the optical 
axis defined by the microscope objective leads to a defocus-dependent tilt. 
Therefore, it is best to introduce in the forward model 
outlined in Sec.~\ref{sec:nlopt}
extra optimization 
parameters to independently adjust these tilts at the BFP for each combination 
of diversities. 
Specifically, steps 4 and 5 in Algorithm~\ref{alg:forward} 
should be reversed in order to apply the polarization diversity before propagating 
to the image plane, and the following additional step should be added after applying the 
polarization diversity:
\begin{enumerate}
    \item[4b.] Apply a phase tilt 
    $T(\bs \ell) = \exp \left(\im 2 \pi \bt t^{(\zeta,p)} \cdot \bs \ell \right)$ 
    to each diversity. \textbf{Optimization parameters:} $\bt t^{(\zeta,p)} =( t_x^{(\zeta,p)} , t_y^{(\zeta,p)})$.
\end{enumerate}
The downside of including these extra parameters is that the number of FFTs needed for each iteration increases by a factor equal to the number of polarization projection steps \red{(see Table.~\ref{tab:run}).
However, the total run time remains quite manageable even if the 2D blurring model is used.}

All the optimization parameters for the retrieval procedure were used, 
except for the bead radius (for the blurring) since it 
can be neglected due to the small bead size ($R_b=10\text{nm}$). 
The results 
of the retrieval process are shown in Fig.~\ref{fig:exp}, where the strong agreement between
the measured PZ-stack and the one modeled with the 
retrieved BDPP \red{(see SM Sec. SVIII for a full comparison)} can be appreciated. 
Moreover, the presence of an SEO is visible 
from the retrieved BDPP. 
In this case, since it is known that there is an SEO at the BFP, 
it is worth trying to separate the total retrieved BDPP into a misaligned ideal SEO and a BDPP 
one containing the scalar and polarization aberrations of the system. 
It should be 
noted that scalar tilts and defocus are not shown as part of the scalar 
aberrations.  From this decomposition it can be seen that the largest birefringent contribution 
comes from the SEO, but the aberrations are not negligible and must be 
taken into account (see SM Sec. SVIII for more details). 
Note also that the polarization aberrations show 
larger variations that the scalar ones, which are almost flat, showing the need to consider polarization aberrations when using polarization-dependent systems. 
\red{Given that for a retrieval from experimental data the ground truth is not known \emph{a priori}, the same retrieval was performed for other beads within the field of view  to corroborate the results. The retrieved BDPPs are indeed highly correlated to the one shown in Fig.~\ref{fig:exp} 
(see SM Sec. SVIII for full details).}

%
%
%
%
%
%
%
%
%
%

\section{Conclusions}

A methodology and phase retrieval algorithm were presented for the characterization of 
birefringent distributions at the pupil plane (BDPP) from stacks of PSFs. 
In particular, it was shown that, for polarization-dependent systems, aberrations 
should be modeled as a BDPP encoding both scalar and polarization 
deviations from the ideal system, and that the use of polarization diversity is 
essential for its proper characterization. The software program \textsc{pyPSFstack} 
created and used for the modeling and retrieval presented in this work is freely available. In particular, the 
BDPP retrieval implementation based on \textsc{PyTorch} makes this software 
flexible and customizable. This is shown in this work by incorporating several 
optimization parameters apart from those used to describe the BDPP, 
as well as the blurring models presented in Ref.~\onlinecite{alemancastaneda2022using}. Even when considering these extra parameters and models the runtime remains manageable. Additionally, this software can also be used for the retrieval of scalar 
BDPPs, as shown in the  SM Sec. SVII for a mask known as the tetrapod.\cite{shechtman2014optimal}

While it was assumed that the sources used for the characterization were 
unpolarized, the retrieval model and algorithm presented here can be easily adapted 
to sources that are dipolar, partially polarized, or with a fixed 
orientation such as molecules fixed on a surface\cite{Hullman:2021} or in DNA origami.\cite{Adamczyk:2022,Hubner:2021} Likewise, this model can also be used 
to find optimal scalar or BDPPs minimizing a particular combination 
of the Cramèr-Rao bounds for the parameters to be extracted from the shape of the PSFs.
Finally, note that there are some modifications that could be implemented 
in the retrieval algorithm that might allow a faster convergence, 
such as the use of stochastic gradient descent, where only a given set of 
measurements are used during each iteration, or spectral initialization.\cite{candes2015phase}



\section*{Supplementary Material}

See the supplementary material for the exact expression of the Green tensor, an analysis of the various criteria used to define the position for the best focus, the blurring models, a detailed comparison between the scalar and vector models, the effect of noise and the number of diversities in the performance of the retrieval, examples using a pixel-based model for the retrieval of birefringent masks, the retrieval of a scalar mask, and further corroboration of the retrieval from experimental data.

\section*{Funding}
R.G.C. acknowledges funding from the Labex WIFI 
(ANR-10-LABX-24, ANR-10-IDEX-0001-02 PSL*). L.A.C. S.B. and M.A.A. acknowledge funding from ANR-21-CE24-0014 and S.B. from ANR-20-CE42-0003.

\section*{Acknowledgments}
R.G.C. acknowledges S.~W.~Paine and S.~M.~Popoff for useful discussions. L.A.C. acknowledges M.~Sison for the experimental advice. The authors also thank T.~G. Brown for supplying the SEO. 

\section*{Disclosures}

The authors declare no conflicts of interest.

\section*{Data Availability Statement}

Data underlying the results presented in this paper 
are available in Ref.~\onlinecite{gutierrez2023repo}.


\section*{References}
\bibliography{microscopy}

\newpage

\end{document}


\newcommand{\pd}[2]{\frac{\partial #1}{\partial #2}} 
\newcommand{\td}[2]{\frac{d #1}{d #2}} 

\newcommand{\bs}{\boldsymbol}
\newcommand{\bt}{\textbf}
\newcommand{\sech}{\text{sech}}
\newcommand{\erfc}{\text{erfc}}
\newcommand{\bse}{\begin{subequations}}
\newcommand{\ese}{\end{subequations}}
\newcommand{\im}{\text{i}}
\newcommand{\ud}[0]{\mathrm{d}}
\newcommand{\norm}[1]{\left\lVert#1\right\rVert}
\newcommand{\op}{\widehat}

\graphicspath{{Figures/},{../Figures/}} 
\allowdisplaybreaks

\renewcommand{\theequation}{S\arabic{equation}} 
\renewcommand{\thefigure}{S\arabic{figure}}
\renewcommand{\thesection}{S\Roman{section}}


\title[Vectorial phase retrieval  -- SM]{Vectorial phase retrieval in super-resolution 
polarization microscopy -- Supplementary Material}
\author{R. Guti\'errez-Cuevas}
\email{rodrigo.gutierrez-cuevas@espci.fr}
\affiliation{Institut Langevin, ESPCI Paris, Université PSL, CNRS, 75005 Paris, France\looseness=-1}
\affiliation{ 
Aix Marseille Univ, CNRS, Centrale Marseille, 
Institut Fresnel, Marseille 13013, France\looseness=-1
}
\author{L.A. Alem\'an-Casta\~neda}%
\affiliation{ 
Aix Marseille Univ, CNRS, Centrale Marseille, 
Institut Fresnel, Marseille 13013, France\looseness=-1
}%
\author{I. Herrera}
\affiliation{ 
Aix Marseille Univ, CNRS, Centrale Marseille, 
Institut Fresnel, Marseille 13013, France\looseness=-1
}
\author{S. Brasselet}
\affiliation{ 
Aix Marseille Univ, CNRS, Centrale Marseille, 
Institut Fresnel, Marseille 13013, France\looseness=-1
}
\author{M.A. Alonso}
\email{miguel.alonso@fresnel.fr}
\affiliation{ 
Aix Marseille Univ, CNRS, Centrale Marseille, 
Institut Fresnel, Marseille 13013, France\looseness=-1
}
\affiliation{The Institute of Optics, University of Rochester, Rochester, NY 14627, USA\looseness=-1}

\begin{abstract}
This document presents supporting material and proofs for the results presented in the main text. 
Section \ref{sec:green} provides the exact expression of the Green tensor.
Section \ref{sec:focus} presents details about the different criteria used to define the best focus.
Section \ref{sec:blurring} discusses the blurring models used to take into account the bead's size. 
Section \ref{sec:svsv} gives a more detailed comparison between the full vectorial phase retrieval algorithm presented in the main text and a scalar version.
Section \ref{sec:noise} examines the effects of noise, as well as the effects of the number and type of diversity measurements on the accuracy of the retrieval. 
Section \ref{sec:pix} presents the results obtained with a pixel-based model and compares them to those based on Zernike polynomials. 
Section \ref{sec:sca} provides an example of the retrieval of a scalar mask.
Section \ref{sec:exp} provides supplementary results that corroborate the retrieved birefringent distribution at the pupil (BDPP) from experimental data presented in the main text. 
\end{abstract}


\maketitle

\section{Expressions for the Green tensor at the back focal plane}
\label{sec:green}

A closed-form for the 
Green tensor at the back-focal plane for a dipolar source placed close to an 
interface can be obtained, as outlined in Ref. \onlinecite{novotny2006principles}. In particular the components of the  
tensor $\mathbbm g$ in Eq.~(1) of the main text are given by
\begin{align}
\mathbbm g(\bt u) =\frac{1}{\sqrt{\gamma_f(u)}} \left( 
\begin{array}{ccc}
\cos^2 \phi \gamma_f(u) \Phi_2 + \sin^2 \phi \Phi_3 & \cos \phi \sin \phi ( \gamma_f(u) \Phi_2 - \Phi_3) & -u \cos \phi \Phi_1  \\
\cos \phi \sin \phi (\gamma_f(u) \Phi_2 - \Phi_3) &  \sin^2 \phi \gamma_f(u)\Phi_2 + \cos^2 \phi \Phi_3 & - u \sin  \phi \Phi_1
\end{array}
\right).
\end{align}
where 
\begin{align}
  \Phi_1(u) = t_{\rm p}(u) \frac{n_f^2 \gamma_f(u) }{n_i^2 \gamma_i(u)}, \qquad
  \Phi_2(u) = t_{\rm p}(u)\frac{n_f}{n_i}, \qquad
  \Phi_3(u) = t_{\rm s}(u) \frac{n_f \gamma_f(u) }{n_i \gamma_i(u)},
\end{align}
with
\begin{align}
  t_{\rm s}(u) =  \frac{2 n_i \gamma_i(u)}{n_i\gamma_i(u) + n_f \gamma_f(u)}, \qquad
  t_{\rm p}(u) =  \frac{2n_i \gamma_i(u)}{n_f \gamma_i(u) + n_i \gamma_f(u) }, 
\end{align}
being the Fresnel coefficients for p and s polarized light, and
\begin{align}
  \gamma_f(u) = \sqrt{1-u^2}, \qquad \text{and} \qquad \gamma_i(u) = \sqrt{1 -\left(\frac{n_f u}{n_i}\right)^2}.
\end{align}

\section{Choosing the best focus location}
\label{sec:focus}

The index mismatch between the embedding medium and the coverslip,
assumed to be the same as that of the immersion liquid of the 
objective, leads to a shift of the paraxial focus and higher-order 
circularly symmetric aberrations, such as spherical. The presence of 
these aberrations makes the choice of the location producing the best
focus subjective to some extent. Several criteria could 
be used, each leading to a different value of the parameter $\alpha$
as defined in Eq.~(3) in the main text. These criteria are based on
compensating the phase terms coming from the propagation of light
from the source to the interface by a defocus term. 
The difference between them defines the wavefront error, which
can be written in terms of $\delta$ and 
$\alpha$ as
\begin{align}
  \Delta W(u) = \lambda n_f  \delta \left[ \frac{1}{n_r} 
  \sqrt{1 - (n_r u)^2 }
- \alpha \sqrt{1 - u^2} \right],
\end{align}
where $n_r = n_f/n_i$ is the relative refractive index between the 
embedding medium with refractive index $n_i$ and the coverslip with
refractive index $n_f$. To simplify the analysis, the radius 
$u_\text{SAF}= 1/n_r$ at which SAF radiation starts appearing will 
be taken as the semi-aperture.  
The wavefront error 
is used to define various criteria for best focus, such as minimizing
the wavefront root-mean square (rms) error (with  $u' = n_r u$)
\begin{align}
  \Delta W_\text{rms}^2 = \frac{1}{\pi}\int_0^{2\pi}\int_0^{1/n_r}
  \left[ \Delta W(u') - \overline{\Delta W} \right]^2 u'\ud u' \ud \phi, 
  \quad \text{where} \quad 
  \overline{\Delta W} =  \frac{1}{\pi}\int_0^{2\pi}\int_0^{1/n_r} 
  \Delta W(u') u'\ud u' \ud \phi,
\end{align}
or the rms spot size
\begin{align}
   \epsilon_\text{rms}^2 = \int_0^{1/n_r} \int_0^{2\pi} 
  \left[ \epsilon_x (u) - \overline{\epsilon_x} \right]^2 u \ud u \ud \phi, 
  \quad \text{where} \quad \epsilon_x = \partial_{u_x}  \Delta W(u).
\end{align}
Note that only one direction is used due to the rotational symmetry of $\Delta W$.

\begin{figure}
  \centering
  \includegraphics[width=0.99\linewidth]{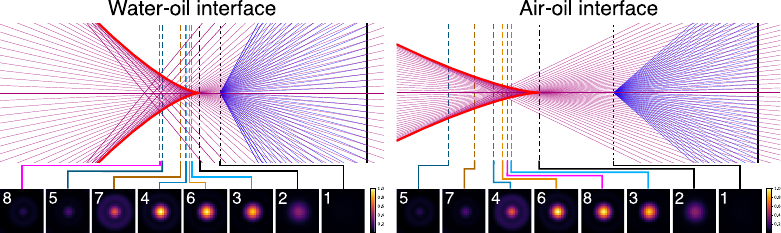}
  \caption{Choosing the location of the best focus. Tracing of the rays
   (blue) emanating from a point source and (purple) refracted 
   by (left) water-oil and (right) air-oil interfaces, 
   marked by a continuous thick black line. By tracing 
   the refracted rays backwards, we see a virtual cusp caustic (red). 
   The position for the various options for best focus are marked as 
   vertical lines, which are connected to the point-spread functions (PSFs) 
   generated by an unpolarized emitter. For the modeling of the PSFs 
   the source was placed at a distance of $|z_0| = 20 \lambda$ in order to 
   enhance the difference between these choices. The numbering on the PSF images corresponds 
   to the one used for the list in the text.}
  \label{fig:sph}
\end{figure}

In the following, several common options for the location of the 
best focus in terms of the parameter $\alpha$ are given:
\begin{enumerate}
  \item the actual location of the source, which leads to $\alpha = 1$;
  \item the paraxial focus, which minimizes the rms wavefront error when it is 
  approximated up to second order which gives $\alpha=n_f/n_i$;
  \item the minimization of the rms wavefront error when it is 
  approximated up to fourth order, which gives
  \begin{align}
    \alpha = \frac{n_r^3 \left(75 n_r^2+19\right)}
    {2 \left(30 n_r^4+15 n_r^2+2\right)};
  \end{align}
  \item the minimization of the rms wavefront error when it is 
  approximated up to sixth order, which gives
  \begin{align}
    \alpha = \frac{n_r^5 \left(36624 n_r^4+9352 n_r^2+4269\right)}{5 \left(5376 n_r^8+2688 n_r^6+1568 n_r^4+336 n_r^2+81\right)};
  \end{align}
  \item the minimization of the rms wavefront error, which gives
  \begin{align}
    \alpha = \frac{7 n_r^3-16 \sqrt{n_r^2-1} n_r^2
    +16 \sqrt{n_r^2-1}+9 \left(n_r^2-1\right)^2 
    \coth ^{-1}(n_r)-9 n_r}{32 n_r^6-32 \sqrt{n_r^2-1} n_r^5
    -48 n_r^4+32 \sqrt{n_r^2-1} n_r^3+12 n_r^2+2};
  \end{align}
  \item the minimization of the rms spot size when the wavefront error is 
  approximated up to fourth order, which gives 
  \begin{align}
    \alpha = \frac{n_r^3 \left(32 n_r^2+11\right)}{24 n_r^4+16 n_r^2+3};
  \end{align}
  \item the minimization of the rms spot size when the wavefront error is 
  approximated up to sixth order, which gives 
  \begin{align}
    \alpha = \frac{n_r^5 \left(1460 n_r^4+512 n_r^2+297\right)}{960 n_r^8+640 n_r^6+480 n_r^4+144 n_r^2+45};
  \end{align}
  \item the circle of least confusion, where the marginal ray intersects 
  the caustic, for which $\alpha$ is given by the solution of the following equation:
  \begin{align}
      \left(\frac{\alpha}{n_r}\right)^{2/3}=\left(\frac{1}{n_r}-\frac{\alpha}{n_r\sqrt{n_r^2+1}}\right)^{2/3}+1.
  \end{align}
\end{enumerate}
As shown in Fig.~\ref{fig:sph}, the PSF generated at the location
of the rms spot size with a fourth-order approximation to the wavefront 
error provides a good starting point and a simple formula. Therefore, this 
value is used as the default in \textsc{pyPSFstack} and in all computations 
performed in the main text. 

The best focus locations just described are derived in an index-matching scenario between the objective immersion liquid and the coverslip, which is common in super-resolution microscopes, and simplifies the calculations since only one interface needs to be considered. 
However, as mentioned in the main manuscript, for setups without index-matching where the coverslip must be considered as a parallel slab of a finite thickness, expressions for the best focal plane can be found in Ref.~\onlinecite{Stallinga:2005}. 
For comparison, note that options 2 and 5, i.e. the ones minimizing the rms wavefront error either in its paraxial approximation or its full expression, correspond respectively to the so-called “Gaussian” and “diffraction” planes in Ref.~\onlinecite{Stallinga:2005}.

\section{Blurring models}
\label{sec:blurring}

\subsection{Exact model}

As a reminder, the PSF of an unpolarized point dipole is given by 
\begin{align}
  I_\text{IP} (\tilde{\bs \rho}) =  \norm{\bt G_\text{IP} }^2 
  = \sum_{i=x,y} \sum_{j=x,y,z} |G_{\text{IP},ij}( \tilde{\bs \rho})|^2.
\end{align}
However, fluorescent nanobeads used for characterizing the performance of a microscope can have a non-negligible size that must be taken into account since it leads to the blurring of the PSFs. As outlined in Ref.~\onlinecite{alemancastaneda2022using}, the blurred PSF generated by a fluorescent bead can be obtained by integrating the PSF of unpolarized point dipoles over its volume
\begin{align}
I_{b} (\tilde{\bs \rho} ; \bt r_0, R_b) = \mathcal{B}\left[  I_\text{IP} (\tilde{\bs \rho} ; \bt r_0); R_b \right] = \iiint  I_\text{IP} (\tilde{\bs \rho} ; \bt r) \crc\left( \frac{\norm{\bt r - \bt r_0}}{R_b} \right) \ud^3 \bt r ,
\end{align}
with $\bt r = \bs \rho + \hat{\bt z} z$, and where  $\bt r_0$ denotes the location of the center of the bead with respect to the optical axis and interface, $R_b$ is the bead's diameter, and $\crc(r) =1$ if $0\leq r \leq 1$ and $0$ otherwise. Note that it was assumed that the bead emits as a transparent sphere in which all points within the volume emit light. Another shell-based emission model was also presented in Ref.~\onlinecite{alemancastaneda2022using}, but is not discussed here since a full sphere model is deemed more adequate for the fluorescent beads used.

Given the three-dimensional nature of the beads, their PSF is not expressible as a 
single convolution since the system is not translation invariant along the $z$ axis. 
Nonetheless, by leveraging the transverse translation invariance 
$I_\text{IP} (\tilde{\bs \rho} ; \bt r_0 + \bs \rho) =I_\text{IP} (\tilde{\bs \rho} - M \bs \rho ; \bt r_0)$ 
and separating the integral into its transverse an longitudinal components, it is possible 
to express it as the integral of a two-dimensional convolution: 
\begin{align}
I_{b} (\tilde{\bs \rho} ; \bt r_0, R_b) 
=&  \int_{z_0-R_b}^{z_0+R_b}\iint  I_i (\tilde{\bs \rho} ; \bs \rho + \hat{\bt z} z) \crc \left( \frac{\rho}{\sqrt{R_b^2 -(z-z_0)^2}} \right)   \ud^2 \tilde{\bs \rho} \ud z  \\
=& \frac{1}{M^2} \int_{-R_b}^{R_b} \iint  I_\text{IP} [\tilde{\bs \rho} -\tilde{\bs \rho}_M;  \hat{\bt z} (z_0+\zeta)] \crc \left[ \frac{\tilde{\rho}_M}{M\sqrt{R_b^2 -\zeta^2}} \right]  \ud^2 \tilde{\bs \rho}_M  \ud \zeta \\
=& \frac{1}{M^2} \int_{-R_b}^{R_b}I_\text{IP} \left[\tilde{\bs \rho};  \hat{\bt z} (z_0+\zeta)\right] \ast
\crc \left[ \frac{\tilde{\rho}}{M\sqrt{R_b^2 -\zeta^2}} \right]  \ud \zeta ,
\end{align}
where $\tilde{\bs \rho}_M = M \bs \rho$ (hence can be seen as a variable at the detector/image plane), the bead is assumed to be located along the optical axis so that $\bt r_0 = \hat{\bt z} z_0$, and we defined $\zeta=z-z_0$. This greatly simplifies the numerical calculation since only PSFs for varying distances $z$ need to be computed. 

Using the convolution theorem, and remembering that the Fourier conjugate variable to the pupil coordinate $\bt u$ is the scaled transverse coordinate $\tilde{\bs \rho}' = n_f \tilde{\bs \rho} / M \lambda$, the convolution can be rewritten as
\begin{align}
I_\text{IP} &\left[\tilde{\bs \rho};  \hat{\bt z} (z_0+\zeta)\right] \ast
\crc \left[ \frac{\tilde{\rho}}{M\sqrt{R_b^2 -\zeta^2}} \right]\\
     =  
    &  \left( \frac{M \lambda}{n_f} \right)^2 \mathcal{F}_{\bt u \rightarrow \tilde{\bs \rho}'}^{-1} \left\{\mathcal{F}_{\tilde{\bs \rho}' \rightarrow \bt u}\left\{ I_\text{IP} \left[\frac{M\lambda}{n_f}\tilde{\bs \rho}';  \hat{\bt z} (z_0+\zeta)\right] \right\}
    K_\text{EM} (\bt u)  \right\},
\end{align}
where the blurring kernel in pupil coordinates is given by
\begin{align}
K_\text{EM} (\bt u) 
    = &  k n_f\sqrt{R_b^2 -\zeta^2} \frac{J_1\left(k u n_f\sqrt{R_b^2 -\zeta^2}\right)}{2 \pi u},
\end{align}
and the Fourier transform is defined as
\begin{align}
\mathcal{F}_{\tilde{\bs \rho}' \rightarrow \bt u}\left\{f(\tilde{\bs \rho}')\right\} = &
\iint f(\tilde{\bs \rho}') e^{\im 2\pi \tilde{\bs \rho}'  \cdot \bt u } \ud^2 \tilde{\bs \rho}' .
\end{align}

\subsection{Semi-analytic model}

Following Ref.~\onlinecite{alemancastaneda2022using}, instead of having to compute the intensity distribution for several values of $\zeta$, i.e.
for several slices of the bead, we can instead perform a Taylor expansion of the intensity 
around $\zeta = 0$, that is the center of the bead,
\begin{align}
    I_\text{IP} [\tilde{\bs \rho} ; \hat{\bt z} (z_0+\zeta)] = \sum_{m=0}^\infty \frac{1}{m!} \zeta^m\pd{^m}{\zeta^m}I_\text{IP}(\tilde{\bs \rho} ; \hat{\bt z} z_0)
\end{align}
The $\zeta$ dependence comes from the phase factor giving rise to the angle fluorescence (SAF) radiation, 
and assuming that $|\zeta|<|z_0|$ (the bead cannot be embedded in the coverslip) then $|z_0+\zeta|=|z_0|+\zeta$ and
\begin{align}
    I_m (\tilde{\bs \rho} ; z_0) =  \pd{^m}{\zeta^m}I_\text{IP}(\tilde{\bs \rho} ; \hat{\bt z} z_0)
    = \sum_{l=0}^m \frac{m!}{l!(m-l)!} \text{Tr}\left[ \bt G_l^\dagger (\tilde{\bs \rho} ; z_0) 
    \cdot \bt G_{m-l} (\tilde{\bs \rho} ; z_0) \right]
\end{align}
where
\begin{align}
\bt G_l (\tilde{\bs \rho} ; z_0) =  \iint \left[ \im k n_i \sqrt{1-\left(\frac{n_f u}{n_i}\right)^2} \right]^l \bt J_\text{M}(\bt u)  
\cdot \bt G_0  (\pm \bt u) e^{- \im k  \frac{n_f \tilde{\bs \rho}}{M}  \cdot \bt u } \ud \bt u .
\end{align}
With this expansion, the intensity distribution of the bead can be written as 
\begin{align}
    I_{b} (\tilde{\bs \rho} ; \bt r_0, R_b) 
=& \frac{1}{M^2} \sum_{m=0}^\infty \frac{1}{m!}\int_{-R_b}^{R_b} \iint  
  \zeta^m I_m (\tilde{\bs \rho} - \tilde{\bs \rho}_M ; z_0)
\crc \left[ \frac{\tilde{\rho}_M}{M\sqrt{R_b^2 -\zeta^2}} \right]  \ud^2 \tilde{\bs \rho}_M  \ud \zeta .
\end{align}
The advantage of this form is that the integral in $\zeta$ can be computed analytically; it vanishes for  odd $m$ while for even $m$ it is given by 
\begin{align}
    \int_{-R_b}^{R_b}\zeta^m\crc \left[ \frac{\tilde{\rho}}{M\sqrt{R_b^2 -\zeta^2}} \right] \ud \zeta
    =& \frac{2}{m+1} \left[ R^2_b - \left(\frac{\tilde{\rho}}{M}\right)^2\right]^{\frac{m+1}{2}}.
\end{align}
Having performed this integral, the intensity distribution produced by the bead can be written as a series of two-dimensional convolutions:
\begin{align}
I_{b} (\tilde{\bs \rho} ; \bt r_0, R_b) 
=& \frac{1}{M^2} \sum_{l=0}^\infty \frac{1}{2l!} \frac{2}{2l+1} \iint \left[ R^2_b - \left(\frac{\tilde{\rho}_M}{M}\right)^2\right]^{\frac{2l+1}{2}}   I_{2l} (\tilde{\bs \rho} - \tilde{\bs \rho}_M ; z_0) \ud^2 \tilde{\bs \rho}_M .
\end{align}
Moreover, the expression of the blurring kernels in the Fourier plane can also be computed analytically
\begin{align}
    K_{2l} (\bt u) 
      =&\frac{1}{2l!} \frac{2}{2l+1} \iint \left[ R^2_b - \left(\frac{\lambda \tilde{\rho}'}{n_f}\right)^2\right]^{\frac{2l+1}{2}}
     e^{\im 2\pi  \tilde{\bs \rho}'  \cdot \bt u } \ud^2 \tilde{\bs \rho}' \\
     =&4\pi \frac{R_b^{2l+3}}{\lambda^2 n_f^2}~\frac{1}{(k n_f R_b u)^{l+1}} j_{l+1}\left(kn_f R_b u\right),
\end{align}
where $j_n$ is a spherical Bessel function of order $n$. 
As discussed in Ref.~\onlinecite{alemancastaneda2022using}, for small size beads, the zeroth order approximation suffices and hence only the first blurring Kernel must be used with the intensity distribution of an unpolarized source at the center of the bead. 
This is the 2D model used in the main text. 
Note that this result differs from other approaches generally used in the literature, such as convolutions with a hard disk\cite{sheng_liu_universal_2023} or a Gaussian\cite{li_global_2022, ferdman2020vipr}.
For the 3D model, it suffices to include some of the next terms in the series expansion.

\section{Scalar VS vectorial model}
\label{sec:svsv}

\begin{figure}
  \centering
  \includegraphics[width=0.9\linewidth]{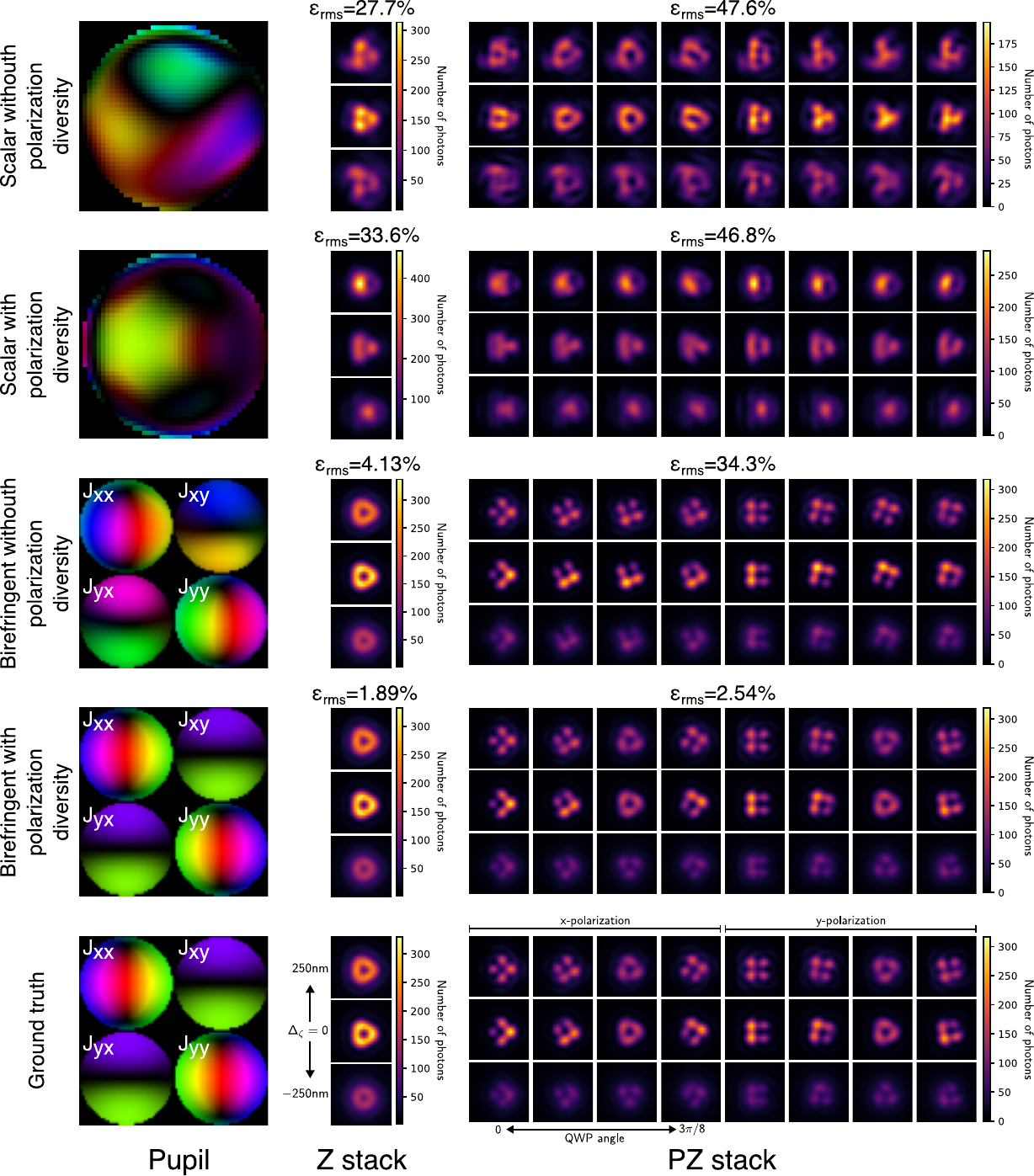}
  \caption{Scalar vs vector phase retrievals. Results of the phase retrieval methods using the scalar (first and second rows) or birefringent (third and fourth rows) models for the BDPP without (first and third rows) and with (second and fourth rows) polarization diversity. The ground truth (fifth row) is shown for comparison.
  For each case, the retrieved BDPP (first column) along with samples of the Z-stack (second column) and PZ-stack (third column) are shown.
  To assess the performance of the retrieval, the rms error, $\varepsilon_\mathrm{rms}$, between the retrieved and true Z-stack and PZ-stack are also reported.}
  \label{fig:svsv}
\end{figure}

In addition to the results of the numerical experiments presented in the main text, here a more in-depth comparison between the scalar and vector models for the phase retrieval is presented. Note that the aberrations and polarization distortions are represented by a birefringent
distribution at the pupil plane (BDPP) modeled with a spatially-varying Jones matrix.
Using the PSF stack generated with an SEO, the four following cases are considered:
\begin{enumerate}
  \item a scalar model for the BDPP without polarization diversity;
  \item a scalar model for the BDPP with polarization diversity;
  \item a birefringent model for the BDPP without polarization diversity;
  \item a birefringent model for the BDPP with polarization diversity.
\end{enumerate}
The results are presented in Fig.~\ref{fig:svsv}. 
Like for the results presented in the main text, the PZ-stacks are obtained by taking images between $-250$nm to $250$nm of the RFP with a step size of 50nm and projecting the polarization by rotating a quarter wave plate (QWP) from $0$ to $3\pi/8$ with a step of $\pi/8$, and then separating in horizontal and vertical linear polarizations. 
The Z-stacks are obtained by taking images between $-250$nm to $250$nm of the RFP with a reduced step size of 5nm. 

The first thing to notice is that a scalar model for the BDPP fails to accurately reproduce the Z- and PZ-stacks generated with the SEO, whether polarization diversity is used or not.  
This shows that a scalar model is generally not sufficient to reproduce the PSFs generated by birefringent widows. 
In comparison, when a fill birefringent model is used the Z-stack can be accurately retrieved whether polarization diversity is used or not, as seen in Fig.~\ref{fig:svsv} (second column) in which both the root-mean square (rms) error, $\varepsilon_\mathrm{rms}$, are comparable. However, in order to reproduce the PZ-stack accurately, i.e. for different polarization projections, polarization diversity is necessary; as shown in  Fig.~\ref{fig:svsv} (third column) the conventional retrieval without polarization diversity fails and hence its overall $\varepsilon_\mathrm{rms}$ is much higher. This failure is not surprising since, without polarization diversity, the phase retrieval algorithm cannot distinguish between unitary transformations of the true BDPP and thus will, most likely, converge to a wrong solution. 

\section{Effects of noise and of number and type of diversity measurements}
\label{sec:noise}

An important choice to be made for the vector phase retrieval algorithm presented in the main text is the number of phase and polarization  diversity measurements used. 
To determine some guidelines, the performance of the retrieval of an SEO is studied as a function of the number of detected photons, which controls the noise, for various combinations of phase and polarization diversity measurements. 
For a given number of detected photons (taken to be the number of incident photons in the brightest pixel), 50 retrievals are performed using PZ-stacks with five phase diversity measurements (obtained by taking images between $-100$nm to $100$nm of the RFP with a step size of 50nm) and with two (quarter-wave plate at $\pi/4$ followed by Wollaston prism), four (quarter-wave plate at $0$ and $\pi/4$ followed by Wollaston prism) and eight (quarter-wave plate at $0, \pi/8, \pi/4$ and $3\pi/8$ followed by Wollaston prism) polarization projection diversity. Similar PZ-stacks are used but with eleven phase diversity measurements (obtained by taking images between $-250$nm to $250$nm of the RFP with a step size of 50nm). 
For every case, the background number of photons is equal to 50 for every pixel. 
To assess the accuracy of the retrieval, two metrics are used: the correlation between the retrieved BDPP and the true one, and the rms error between the retrieved and true PSFs generated by point dipoles along each of the three Cartesian axes plus an unpolarized one (as those shown in Fig.~3 in the main text). 

\begin{figure}
  \centering
  \includegraphics[width=0.9\linewidth]{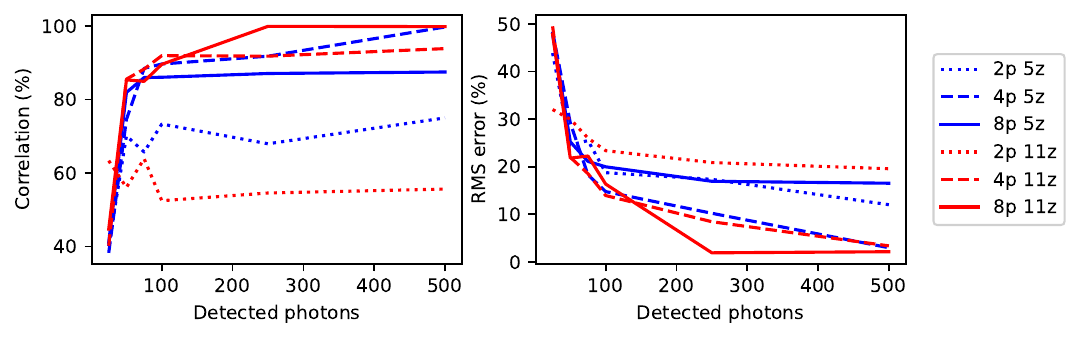}
  \caption{Performance of the retrieval process as a function of the number of detected photons for various combinations of phase and polarization diversity measurements. Both the mean correlation (left) between the retrieved and the true BDPP and the rms error (right) between the retrieved and true PSFs generated by point dipoles along each of the three Cartesian axes and an unpolarized dipole are shown for six different combinations of phase and polarization diversity measurements.
  The phase and polarization diversities are labeled $N_p\text p$ and $N_z\text z$, where $N_p$ and $N_z$ are the number of measurements used comprising each diversity respectively. 
  Each point is obtained by averaging the results of 50 retrieval runs using the PZ stack generated by an SEO. The reported detected photon number represent the maximum number of photons incident in the brightest pixel and the background number of photons is equal to 50 for every pixel. }
  \label{fig:noise}
\end{figure}

The results are shown in Fig.~\ref{fig:noise}, where similar trends can be appreciated for both metrics. 
The first thing to notice is that a diversity of just two polarization projections is not sufficient to fully determine a BDPP, no matter how many photons are detected. 
With just two polarization projections not all the elements of the Jones matrix can be determined, and thus the algorithm almost always converges to the wrong BDPP.
When the number of polarization projections is increased to four, the performance improves significantly, and accuracy increases with the number of detected photons. 
However, a successful retrieval is not guaranteed. 
We observe that by using an eight polarization projection diversity and eleven phase diversity measurements is a successful retrieval achieved when the number of incident photons in the brightest pixel is larger than 250, which is the case for the results presented in the main text. 
Therefore, an increase in the number of diversity measurements and of incident photons generally improves the performance of the algorithm.

\section{Pixel-based model}
\label{sec:pix}

\begin{figure}
  \centering
  \includegraphics[width=0.9\linewidth]{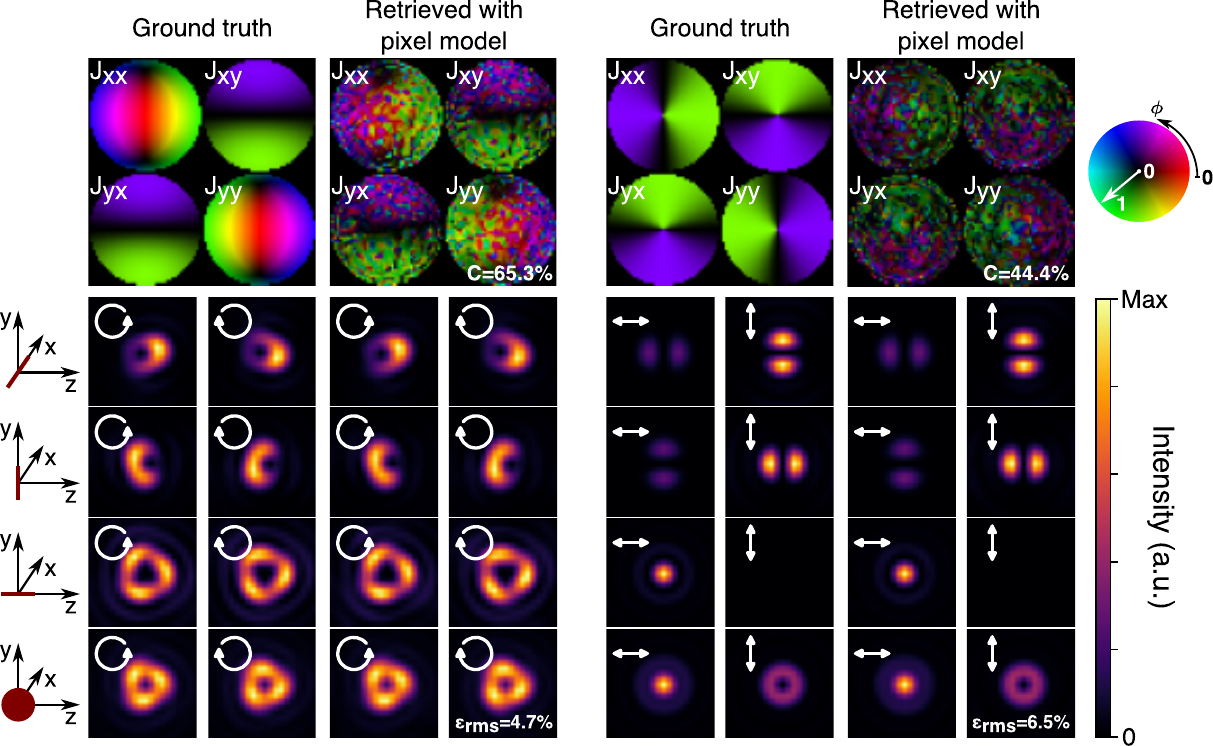}
  \caption{Pixel-based retrieval.
  (first row) Elements of the Jones matrix for the ground truth and the retrieved BDPPs with a pixel-based model for (left) an SEO and (right) a q-plate. 
  Also shown are the PSFs generated by point dipoles (second to fourth rows) oriented along each of the three Cartesian axes, and (last row) an unpolarized dipole, for each of the standard projectors used for each type of birenfringent window. 
  For the SEO the PSFs are projected onto left- and right-circular polarizations, while for the q-plate they are projected onto linear horizontal and vertical polarizations.
  Also indicated are the correlation between the retrieved and true BDPP and the rms error  between the retrieved and true PSFs generated by point dipoles along each of the three Cartesian axes, and an unpolarized dipole are also reported.}
  \label{fig:pix}
\end{figure}

When using a pixel-based model for the BDPP, it is still convenient to use a decomposition as the one in Eq.~(4) of the main text. Then, instead of 
expanding the components $W$ and $q_j$ into Zernike polynomials, these are 
discretized using the same sampling outline for the numerical 
implementation, and each value labeled by the lexicographic index $\bs \ell$ 
becomes an optimization parameter with all those falling outside the 
aperture being set to zero. A technical point for this model is that, for the 
optimization, the defocus parameter $\Delta$ should be removed since it will 
be automatically taken care of. The defocus term 
can then be removed from the retrieved BDPP by fitting to it the appropriate function.

\begin{figure}
  \centering
  \includegraphics[width=0.7\linewidth]{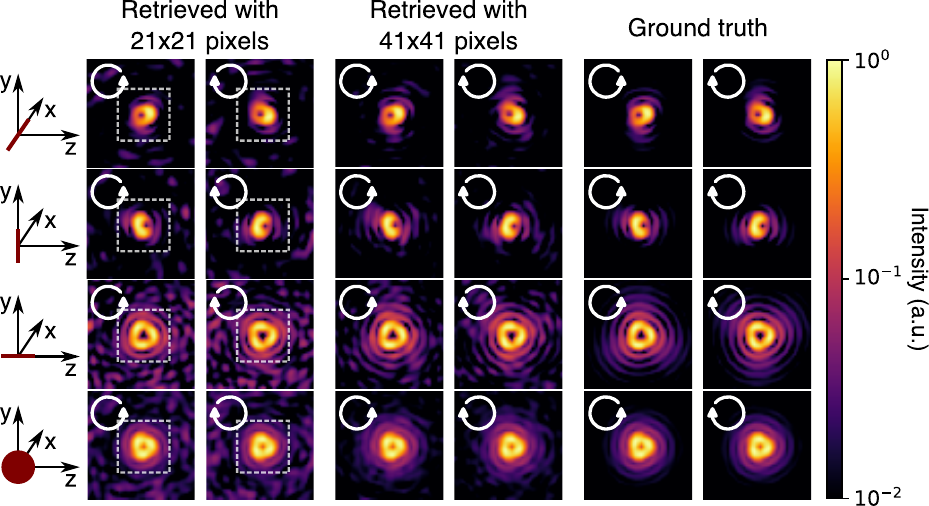}
  \caption{PSFs over a larger area in log scale. 
  PSFs generated by point dipoles (second to fourth rows) oriented along each of the three Cartesian axes and (last row) an unpolarized dipole, plotted over a larger area (41$\times$41 pixels instead of 21$\times$21 pixels). 
  The PSFs are those obtained when using the pixel-based model, for an area of 21$\times$21 pixels (first column), and 41$\times$41 pixels  (second column). These results are compared to the ground truth (third column). 
  The dashed squares in the first column denote the area used for the retrieval, namely 21$\times$21 pixels.}
  \label{fig:gran}
\end{figure}

This discretized model is also implemented in \textsc{pyPSFstack}, and 
Fig.~\ref{fig:pix} shows the results obtained when it is used for the 
retrieval of the SEO and q-plate with the same parameters as those used to 
obtain the results presented in Fig.~3 in the main text. While the general 
shape of the true BDPPs can be identified from the retrieved ones, there is 
a distinctive granularity with fast variation from on pixel to the next. 
This granularity decreases the correlation between the retrieved BDPP and the ground truth, but does not affect the modeling of the PSFs which still present a small error (see the rms errors reported in Fig.~\ref{fig:pix}).
The granularity of the retrieved BDPP is a consequence of the increased number of parameters used to model the BDPP, which makes the optimization prone to fall into local minima. 
Nonetheless, these fast variations in the BDPP have little to no impact on the resulting PSFs since they mostly send light outside the regions used to compute the cost 
function, which are controlled by $w$ in Eq.~(14) of the main text.
This is shown in the first column of Fig.~\ref{fig:gran} where the same PSFs shown in Fig.~\ref{fig:pix} for the SEO are plotted over a larger area and in log scale. 
The dashed square represents the area used for the retrieval and is the one used to compute the rms error in Fig.~\ref{fig:pix}.
Even in log scale it is hard to see any the difference between the retrieved and true PSFs lying inside the dashed square. 
The largest differences happen outside, where a speckle pattern is formed by the rapid fluctuations of the retrieved BDPP.
Therefore, these retrieved BDPPs effectively behave as low-pass 
filtered versions of themselves within the region used for the retrieval. One way to obtain a BDPP that is closer to the true one is then to filter out the high frequencies.
Another is to use a larger area for the retrieval which results in the PSFs shown in the second column of Fig.~\ref{fig:gran}.

\begin{figure}
  \centering
  \includegraphics[width=0.9\linewidth]{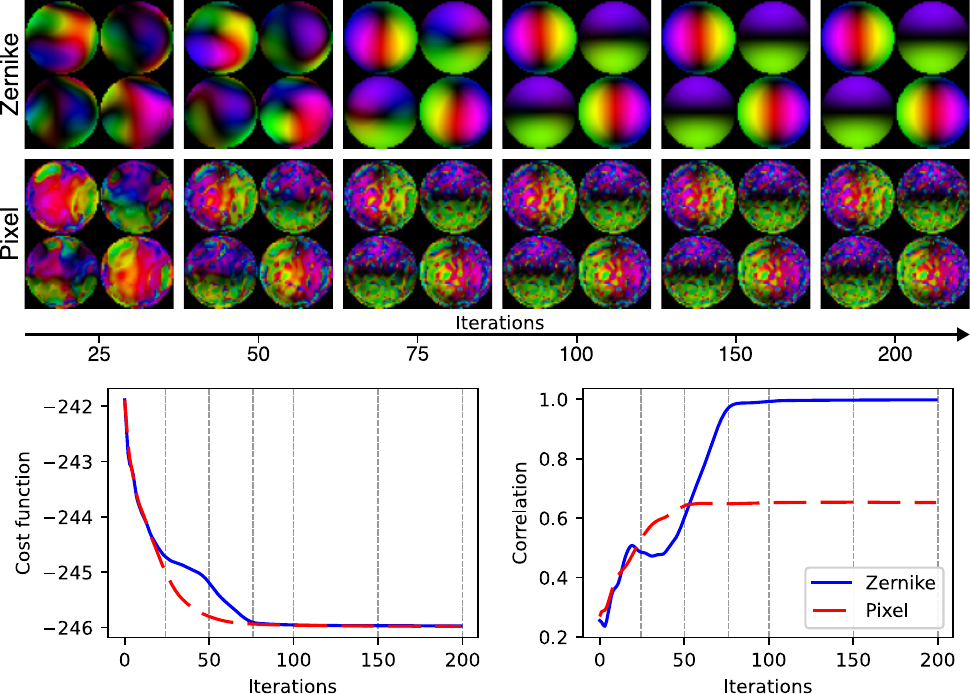}
  \caption{Evolution of the estimation process of the vector phase retrieval algorithm.
  (first row) Evolution of the retrieved BDPP with respect to the iteration for both the Zernike and pixel-based models. 
  (second row) Evolution of the cost function $\mathcal{C}$ and the correlation between the retrieved BDPP and the true one with respect to the iteration.}
  \label{fig:pvsz}
\end{figure}

Let us now consider the convergence of the algorithm and the evolution of the BDPP for both the Zernike and pixel models. 
Figure \ref{fig:pvsz} shows the evolution of the BDPPs, the cost function and the correlation between the retrieved BDPP and the true one (an SEO) with respect to the iteration number. 
Both models converge approximately at the same rate and produce similar values for the cost function.  
However, the granularity of the pixel-based pupil prevents the correlation from approaching unity. 
Moreover, from the evolution of this pupil, it can be seen that the granularity appears from the beginning. 
Therefore, early stopping would not solve this issue, which it is not due to over-fitting but to an increase of the number of local minima.

\section{Retrieval of a scalar phase mask}
\label{sec:sca}

\begin{figure}
  \centering
  \includegraphics[width=0.8\linewidth]{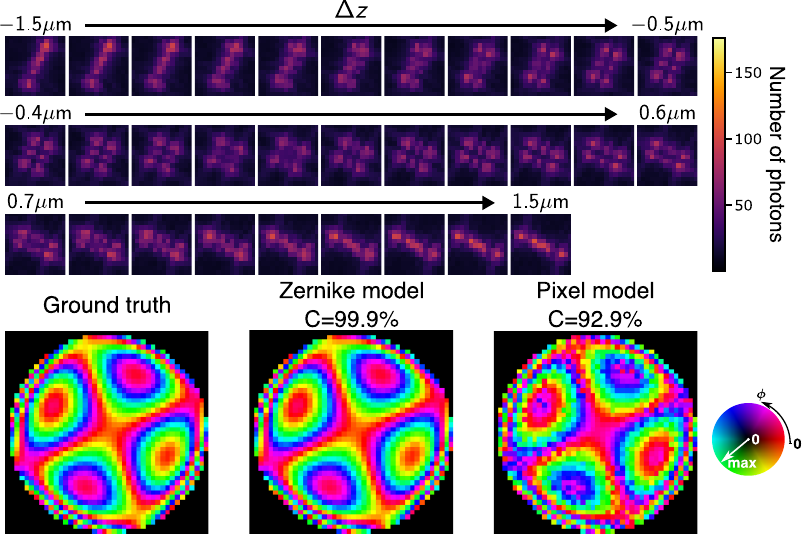}
  \caption{Retrieval of a scalar phase mask.
  (first row) Z-stack of  PSFs modelled with the software \textsc{pyPSFstack} for the tetrapod scalar phase mask. (second row) Scalar pupils for the ground truth and the retrieved pupils with Zernike- and pixel-based models.}
  \label{fig:sca}
\end{figure}

\begin{figure}
  \centering
  \includegraphics[width=0.95\linewidth]{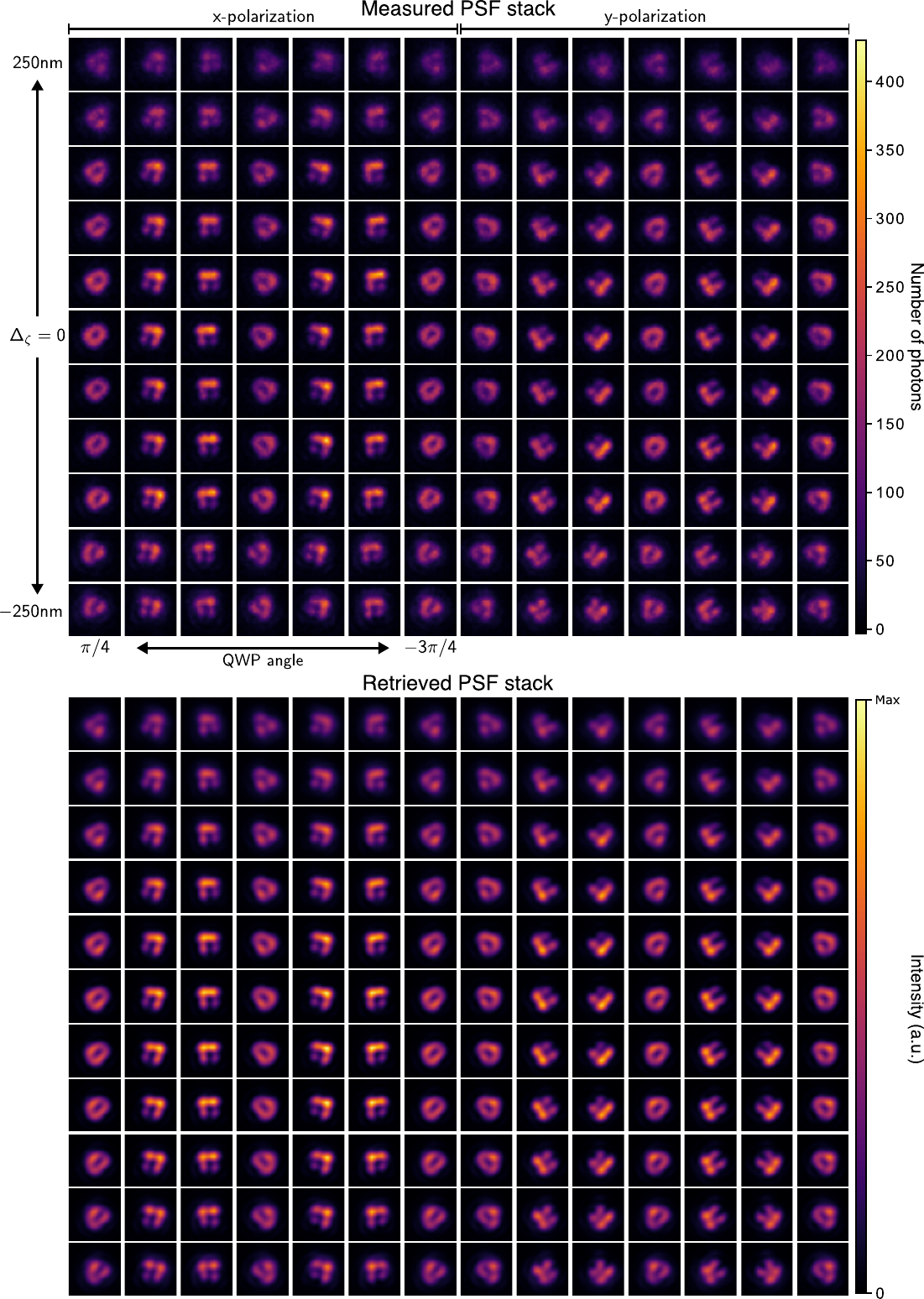}
  \caption{Comparison between the complete measured PZ stack generated with an SEO and the retrieved one. The corresponding retrieved pupil is shown in Fig.~5 of the main text and is the one with the label $0$ in Fig.~\ref{fig:pupils}.}
  \label{fig:expstack}
\end{figure}

As mentioned in the main text, \textsc{pyPSFstack} can also be used for the retrieval of scalar phase distributions at the pupil using both Zernike- and pixel-based models, like those mentioned for BDPPs when $q_j=0$ for $j=1,2,3$. 
Nonetheless, these have been implemented as separate models. 
As an example, the retrieval of a tetrapod phase mask \cite{shechtman2014optimal} is considered. 
Figure~\ref{fig:sca} shows a phase mask designed to optimize the localization for defocus ranging from $-1.5\mu$m to $1.5\mu$m and the Z-stack used for the retrieval. 
Also shown are the results of the retrieval with both the Zernike and pixel-based models, along with the Z-stacks modeled with the retrieved BDPPs. 
We see that both models provide accurate results, with the slight difference of obtaining a smoother BDPP for the Zernike-based model.  
Additionally, it should be mentioned that the runtime for both models is quite similar when using the same sampling since, as mentioned in the main manuscript, the main bottleneck is the number of FFTs required, which is the same for both models, and the gradients for all parameters are computed analytically. 

\section{Verification of experimental retrieval}
\label{sec:exp}

\begin{figure}
  \centering
  \includegraphics[width=0.6\linewidth]{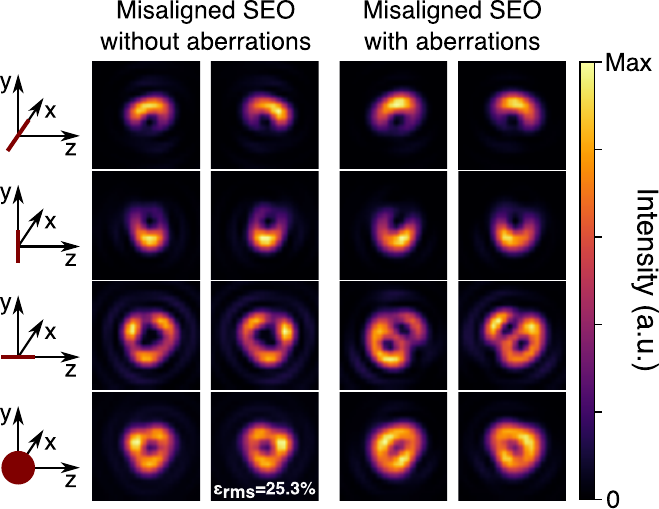}
  \caption{Retrieved BDPPs with and without aberrations. Comparison between the PSFs generated by point dipoles along each of the three Cartesian axes and an unpolarized one, when only a misaligned SEO is used and when the SEO plus the full polarization aberrations are considered. The rms error between the two is also reported.}
  \label{fig:avsna}
\end{figure}

To further support the results of the retrieval of a BDPP from experimental measurements, Fig.~\ref{fig:expstack} compares the full measured PZ- stack to the one generated using the retrieved BDPP. 
Excellent agreement between the two can be appreciated across all the polarization and phase diversity measurements. 
Moreover, to show the need for a proper characterization of the system for single molecule orientation localization microscopy Fig.~\ref{fig:avsna} compares the PSFs generated by point dipoles along each of the three Cartesian axes as well as an unpolarized dipole, when only a misaligned SEO is used and when the SEO plus the full polarization aberrations are considered. 
From their shape and the rms error reported it is clear that without the full polarization characterization the appropriate PSFs cannot be reproduced. 
Additionally,  
from the values retrieved from the algorithm it seems that photobleaching must be taken
into account since there are significant variations in the number of photons emitted by a single bead. However, these numbers of photons do not follow a downward trend
with respect the acquisition time but rather they oscillate. This behavior can be explained by the fact that the
beads contain many fluorophores so that only some of them are photobleached at a given time.

\begin{figure}
  \centering
  \includegraphics[width=0.9\linewidth]{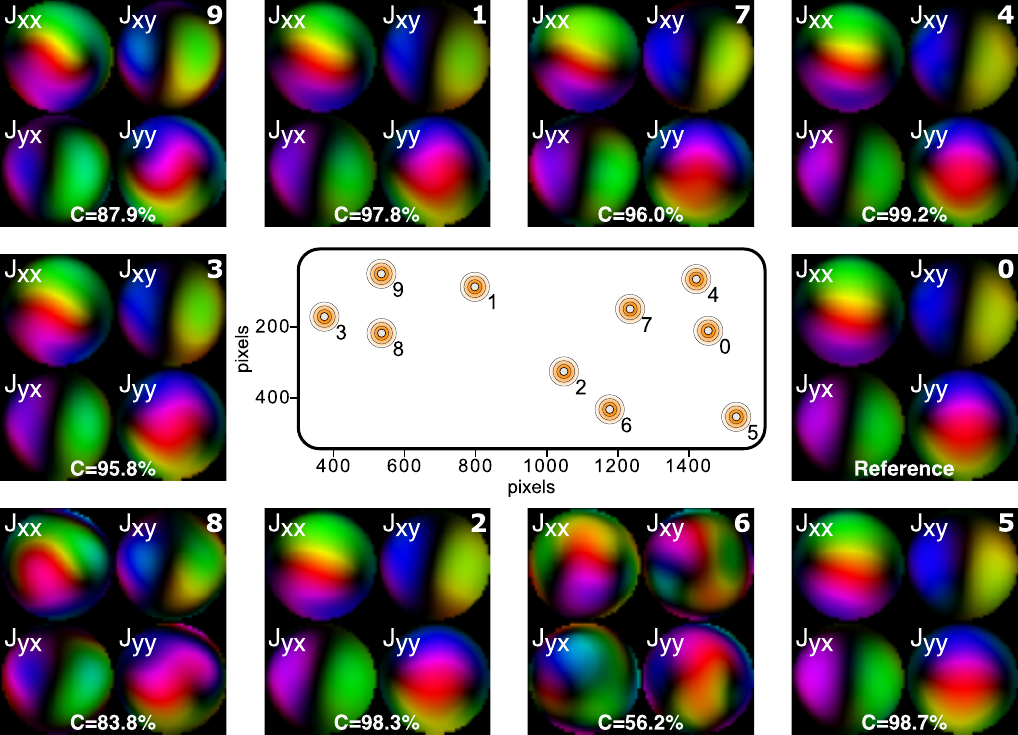}
  \caption{Retrieved BDPPs for various beads across the field of view. The positions of the different beads are shown at the center. The bead  0 is used as a reference and corresponds to the one used for the results presented in the main text in Fig.~\ref{fig:expstack}. All the correlation shown at the bottom of each BDPP are computed with respect to the reference.}
  \label{fig:pupils}
\end{figure}

When retrieving a BDPP from experimental measurements, the ground truth is not known. Therefore, in order to corroborate these results the same retrieval is carried out using the PZ-stack generated by nine more beads positioned across the field of view.
The retrieved BDPPs are then compared to the reference one, shown in the main text. 
The bead leading to the reference BDPP produced the brightest PZ-stack which increases the confidence in the retrieval. 
Moreover, as can be seen from the results presented in Fig.~\ref{fig:pupils}, most PZ-stacks lead to BDPPs that are highly correlated to the reference one. 
Note that some variations are expected due to field-dependent aberrations. 
Only the BDPP retrieved from bead $6$ significantly differs from the reference one, but its PZ stack has less than a third of the photons of those for the reference, impacting the accuracy of the retrieval as shown in Fig.~\ref{fig:noise}.
Nonetheless, if the phase retrieval algorithm is run again, this time setting as a starting point the reference BDPP then it converges to a BDPP, with a lower cost function and with a correlation of 91.5\%.

\newpage
\section*{References}
\bibliography{microscopy}